\documentclass[conference]{IEEEtran}
\IEEEoverridecommandlockouts
\usepackage{cite}
\usepackage{amsmath,amssymb,amsfonts}
\usepackage{algorithmic}
\usepackage{graphicx}
\usepackage{textcomp}
\usepackage{indentfirst}
\usepackage{xcolor}
\usepackage{booktabs}
\usepackage{url}
\usepackage{booktabs}
\usepackage{multirow}
\usepackage{fontenc}
\usepackage{pifont}
\def\BibTeX{{\rm B\kern-.05em{\sc i\kern-.025em b}\kern-.08em
    T\kern-.1667em\lower.7ex\hbox{E}\kern-.125emX}}

\newif\ifdebug
\debugtrue

\newif\ifspec
\specfalse

\makeatletter
\newcommand{\linebreakand}{%
  \end{@IEEEauthorhalign}
  \hfill\mbox{}\par
  \mbox{}\hfill\begin{@IEEEauthorhalign}
}
\makeatother

\newcommand{\spec}[1]{
\ifspec
\textcolor{blue}
{\mbox{}\\
\noindent
\textbf{\underline{Outline of Proposed Content}: {#1}}
}
\fi
}
    
\begin{document}

\title{Large Language Models for Software Engineering: Survey and Open Problems\\
}

\author{
\IEEEauthorblockN{Angela Fan}
\IEEEauthorblockA{\textit{Generative AI Team} \\
\textit{Meta Platforms Inc.}\\
New York, NY, USA}
\and
\IEEEauthorblockN{Beliz Gokkaya}
\IEEEauthorblockA{\textit{PyTorch Team} \\
\textit{Meta Platforms Inc.}\\
Menlo Park, CA, USA}
\and
\IEEEauthorblockN{Mark Harman}
\IEEEauthorblockA{\textit{Instagram Product Foundation} \\
\textit{Meta Platforms Inc.}\\
London, UK}
\linebreakand
\IEEEauthorblockN{Mitya Lyubarskiy}
\IEEEauthorblockA{\textit{Developer Infrastructure} \\
\textit{Meta Platforms Inc.}\\
London, UK}
\and
\IEEEauthorblockN{Shubho Sengupta}
\IEEEauthorblockA{\textit{FAIR} \\
\textit{Meta Platforms Inc.}\\
Menlo Park, CA, USA}
\and
\IEEEauthorblockN{Shin Yoo}
\IEEEauthorblockA{\textit{School of Computing}\\
\textit{KAIST} \\
Daejeon, Korea}
\and
\IEEEauthorblockN{Jie M. Zhang}
\IEEEauthorblockA{\textit{Department of Informatics} \\
\textit{King's College London}\\
London, UK}
}
\maketitle
\thispagestyle{plain}
\pagestyle{plain}

\begin{abstract}
This paper provides a survey of the emerging area of 
Large Language Models (LLMs) for Software Engineering (SE). 
It also sets out open research challenges for the application of LLMs to technical problems faced by software engineers.
LLMs' emergent properties  bring novelty and creativity with applications right across the spectrum of Software Engineering activities including coding, design, requirements, repair, refactoring, performance improvement, documentation and analytics.
However, these very same emergent properties also pose significant technical challenges; we need techniques that can reliably weed out incorrect solutions, such as hallucinations.
Our survey reveals the pivotal role that hybrid techniques (traditional SE plus LLMs) have to play in the development and deployment of reliable, efficient and effective LLM-based SE. 
\end{abstract}

\begin{IEEEkeywords}
Automated Program Repair, 
Documentation generation,
Generative AI, 
Genetic Improvement,
Human-Computer Interaction,
Large Language Models,
Refactoring,
Requirements engineering,
Search Based Software Engineering (SBSE),
Software Analytics,
Software Engineering Education,
Software Processes,
Software Maintenance and Evolution,
Software Testing.

\end{IEEEkeywords}

\section{Introduction}
\spec{this section will introduce the history of artificial intelligence as used by software engineers to tackle technical issues. It will not cover any of the controversial aspects of artificial intelligence, but merely their application to the generation and testing of software artefacts. The focus will be on the scientific research literature, in order to identify the underlying themes that will be developed in the paper, mentioned in the abstract.}

This paper surveys the recent developments, advances and empirical results on LLM-based SE; the application of Large Language Models (LLMs) to Software Engineering (SE) applications.
We use the survey to highlight gaps in this rapidly developing, but as yet embryonic, research literature.
Based on gaps in the literature and technical opportunities, we also identify open problems and challenges for the software engineering research community.

While any survey of such a rapidly expanding area can neither aspire nor claim to be comprehensive, we hope that this survey will provide a useful and relatively complete snapshot of the early universe of this exciting new subdiscipline of Software Engineering: LLM-based Software Engineering.
Although the scientific and technical structure of the field is still emerging, it is already possible to identify trends, productive avenues for future research, and important technical challenges that need to be addressed.

In particular, we are already able to discern important connections to (and resonance with) existing trends and well-established approaches and subdisciplines within Software Engineering.
Furthermore, although we find considerable grounds for optimism, 
there remain important technical challenges, which are likely to inform the research agenda for several years.
Many authors have highlighted, both scientifically and anecdotally, that hallucination is a pervasive problem for LLMs \cite{yang_harnessing_2023} and also that it poses specific problems for LLM-based SE \cite{ma:scope}.
As with human intelligence, hallucination means that the LLM can create fictitious output.
In the context of software engineering, it means that the engineering artefacts created could be incorrect, yet appear plausible; LLMs may introduce bugs.

However, unlike many other applications of LLMs, software engineers are typically blessed with automatable ground truth (software execution), against which most software engineering artefacts can be evaluated.
Also, the software engineering research community has already devoted a great deal of time to producing automated and semi-automated techniques for checking the potentially incorrect results produced by humans.
This means that, for the discipline and the research community, there is a great deal of experience and expertise on which to draw, when tackling the challenges posed by issues like hallucination.

Clearly, automated testing techniques \cite{anandetal:orchestrated,cadar:three-decades,mh:icst15-keynote} will have a central role to play in ensuring correctness, just as they already do for human-engineered artefacts.
When generating entirely new features and systems, automated test data generation suffers from the lack of
an automatable oracle \cite{ebetal:oracle} (an automated technique for determining whether output behaviour is correct for a given input stimulus). 
Given  LLMs' propensity to hallucinate, the Oracle Problem will remain highly relevant, and solutions to it will become all the more impactful~\cite{shin2023assessing}.

However, some SE applications  concern adaption, improvement and development of {\em existing} software systems, for which there {\em is} a readily-available automatable oracle: the functional behaviour of the original system.

In this paper, we call this the `Automated Regression Oracle', an approach that has already proved advantageous in the field of Genetic Improvement \cite{Petke:gisurvey}.
The Automated Regression Oracle simply uses the existing version of the software system as a reference against which to benchmark output from any subsequent adaptions and changes.

Of course, there is a risk of `baking in'  functional incorrectness, since the Automated Regression Oracle cannot detect what the system {\em should} do, but only capture what it currently does.
Therefore, the Automated Regression Oracle can test only  for functional  regressions 
so it is best suited to use cases where the existing functionality is to be maintained.
For example, for non-functional improvements such as performance optimisation and for semantics-preserving refactoring.

The input provided to an LLM will be a natural focus of growing research, and we can expect a rapid development of the literature on prompt engineering and prompt optimisation  \cite{cantino:prompt-engineering-blog}. 
In this survey, we highlight existing work and open challenges for prompt engineering with regard to several specific aspects of software engineering.

The output from an LLM need not be confined purely to code, but can also include other software engineering artefacts, such as requirements, test cases, design diagrams, and documentation. 
In general, the language-based nature of an LLM, allows it to generate any linguistically-defined software engineering artefact.

We typically think of the software engineering artefact as the primary output of the LLM, but it is not the only output.
The {\em explanation} provided with the primary output is also an important output of any LLM.
Our survey highlights the need for much more research, not only into optimising prompt engineering (which focuses on the input to the LLM) but also the need for work on the optimisation of explanations provided with the primary output.

LLMs are inherently nondeterministic:  the same prompt produces different answers on different inference executions (unless the temperature is set to zero, which has often been found to be suboptimal over multiple executions)~\cite{ouyang2023llm}. 
Furthermore, irrespective of the temperature setting, subtle changes in the prompt can lead to very different outputs~\cite{ouyang2023llm}.
As well as motivating
`prompt engineering' and output processing, this nondeterministic behaviour raises challenges for the scientific evaluation of LLM-based Software Engineering:

\begin{quote}
    If results can vary each time we run the process, how can we determine whether a proposed technique achieves an advance over the state of the art?
\end{quote}

This is a problem that has already been well studied in the context of Empirical Software Engineering \cite{easterbrook:selecting} and Search Based Software Engineering (SBSE) \cite{mhamyz:acm-surveys}.
In particular, SBSE bears many similarities to LLM-based Software Engineering, sharing with it the need to achieve robust scientific evaluation in the presence of noisy, non-deterministic, and incomplete results~\cite{arcuri:practical,mhetal:sbse-tutorial}.
There is, therefore, already a mature software engineering literature on just the kind of robust scientific evaluation techniques
needed to cater for LLM-based scientific evaluation. 

For example, well-studied techniques, such as parametric and non-parametric inferential statistics, are now routinely used to provide robust scientific conclusions in the presence of highly non-deterministic algorithms in the SBSE discipline.

\begin{figure*}[h!]
\centerline{\includegraphics[width=0.9\linewidth]{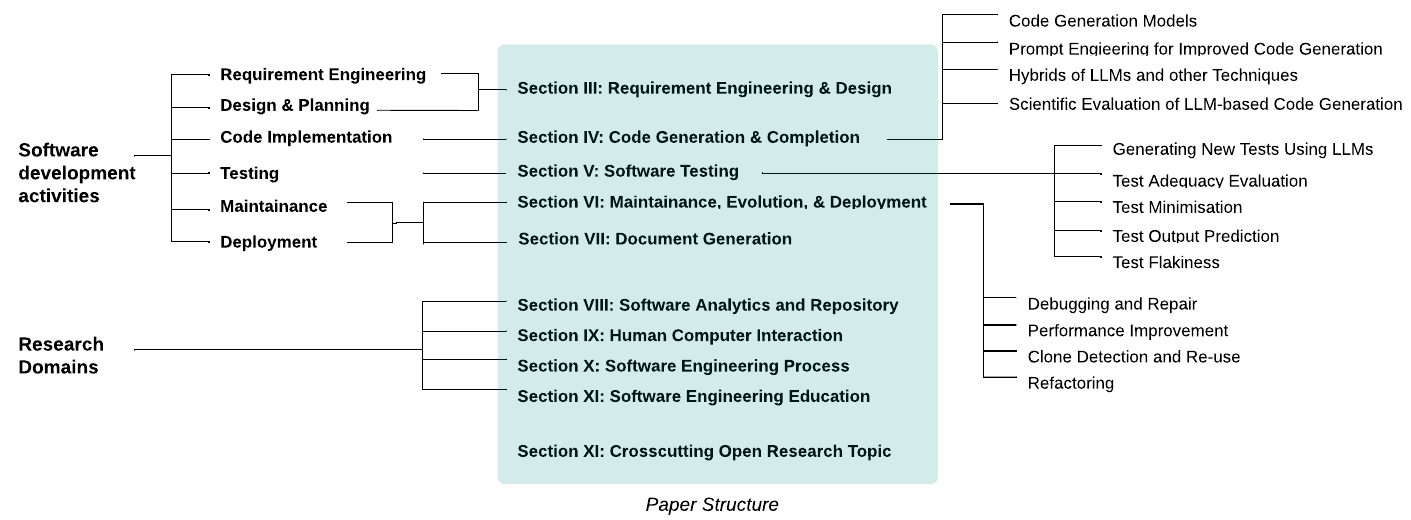}}
\vspace{-3mm}
\caption{A mapping between software development activities, research domains, and the paper structure
\label{fig:structure}}
\end{figure*}


\begin{figure}[ht]
    \centering
    \includegraphics[width=90mm]{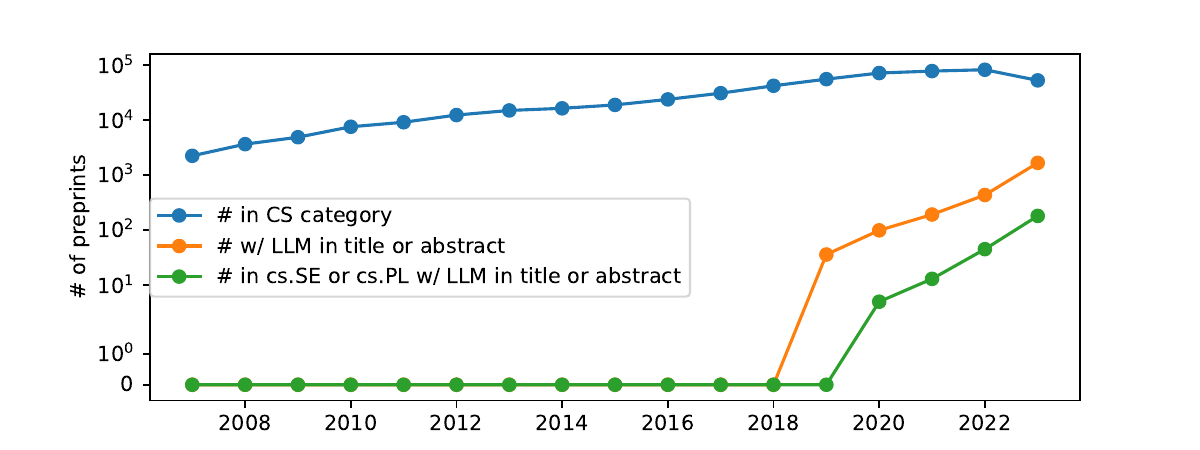}
    \vspace{-5mm}
    \caption{Trends in number of arXiv preprints. 
    The blue line denotes the number of preprints categorised under ``CS''. 
    The orange line denotes the number of preprints in AI (cs.AI), Machine Learning (cs.LG), Neural and Evolutionary Computing (cs.NE), Software Engineering (cs.SE), and Programming Language (cs.PL) whose title or abstract contains either ``Large Language Model'', ``LLM'', or ``GPT''. 
    The green line denotes the number of preprints in SE and PL categories whose title or abstract contains either ``Large Language Model'', ``LLM'', or ``GPT''}
    \label{fig:trend-1}
\end{figure}

\begin{figure}[ht]
    \centering
    \includegraphics[width=90mm]{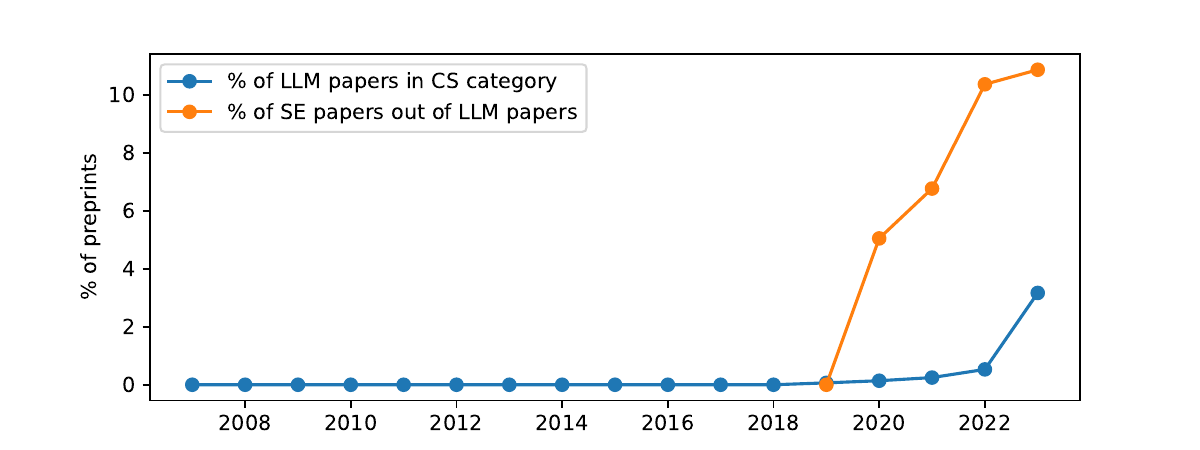}
    \vspace{-5mm}
    \caption{Proportions of LLM papers and SE papers about LLMs. By ``about LLMs'', we mean that either the title or the abstract of a preprint contains ``LLM'', ``Large Language Model'', or ``GPT''.    
    The blue line denotes the percentage of the number of preprints about LLMs out of the number of all preprints in the CS category. The orange line denotes the percentage of the number of preprints about LLMs in cs.SE and cs.PL categories out of all preprints about LLMs}
    \label{fig:trend-2}
\end{figure}

\begin{table}[h]
\centering
\caption{$A$ (ALL) denotes all preprints that are categorised under CS (Computer Science). $L$ (LLM) denotes preprints whose title or abstract includes ``LLM'', ``Large Language Model'', or ``GPT''. $L \cap S$ denotes preprints in cs.SE or cs.PL category whose title or abstract includes the same keywords. Note that the year 2023 only includes data up to 27 July 2023. \label{tab:trend_raw}}
\begin{tabular}{lrrrrr}
\toprule
Year & $|A|$ & $|L|$ & $|L \cap S|$ & $\frac{|L|}{|A|}$(\%) & $\frac{|L\cap S|}{|L|}$(\%) \\
\midrule
2007 &  2,238 &     0 &   0 & 0.00 & 0.00 \\
2008 &  3,645 &     0 &   0 & 0.00 & 0.00 \\
2009 &  4,873 &     0 &   0 & 0.00 & 0.00 \\
2010 &  7,543 &     0 &   0 & 0.00 & 0.00 \\
2011 &  9,114 &     0 &   0 & 0.00 & 0.00 \\
2012 & 12,316 &     0 &   0 & 0.00 & 0.00 \\
2013 & 14,933 &     0 &   0 & 0.00 & 0.00 \\
2014 & 16,320 &     0 &   0 & 0.00 & 0.00 \\
2015 & 18,818 &     0 &   0 & 0.00 & 0.00 \\
2016 & 23,707 &     0 &   0 & 0.00 & 0.00 \\
2017 & 30,746 &     0 &   0 & 0.00 &  0.00 \\
2018 & 41,927 &     0 &   0 & 0.00 &  0.00 \\
2019 & 55,325 &    36 &   0 & 0.00 &  0.00 \\
2020 & 71,431 &    99 &   5 & 0.00 &  5.05 \\
2021 & 77,520 &   192 &  13 & 0.25 &  6.77 \\
2022 & 81,964 &   434 &  45 & 0.53 & 10.36 \\
2023 & 52,547 & 1,665 & 181 & 3.17 & 10.87 \\
\bottomrule
\end{tabular}
\end{table}

In order to understand the growth trends within LLM-based Software Engineering,
we performed a manual analysis of data on the number of publications on specific topics from arXiv.
Table~\ref{tab:trend_raw} contains the raw data\footnote{The numbers for 2023 are underestimated since the data was accessed in July 2023.}, which was manually extracted from the arXiv metadata dump made publicly available via Kaggle (\url{https://www.kaggle.com/datasets/Cornell-University/arxiv}), accessed on the $27^{th.}$ July 2023. 
We first filtered out publications for which the classification code does not start with the \texttt{cs} prefix (i.e., Computer Science), resulting in column $A$. 

To identify Computer Science papers that are relevant to LLMs, 
we filtered the publications into subcategories on 
artificial intelligence (cs.AI), 
machine learning (cs.LG), 
neural and evolutionary computation (cs.NE), 
software engineering (cs.SE), 
and programming language (cs.PL) 
using the queries ``Large Language Model'', ``LLM'', and ``GPT'' in either the title or the abstract (we manually excluded instances of overloaded acronyms such as GPT for General Planning Tool), 
resulting in column $L$. 
Finally, we used the same queries to identify LLM-based Software Engineering papers in software engineering (cs.SE) and programming language (cs.PL). 
These queries are inherently approximate, so we confine ourselves only to conclusions based on overall trends for which there is strong evidence rather than specific details of the numbers observed.
Nevertheless, we report the raw numbers observed to support replication by others.

Figure~\ref{fig:trend-1}, shows the growth in the  number of arXiv-published papers on Computer Science ($|A|$, in Blue), and on LLMs ($|L|$, in orange). 
Those papers specifically on Software Engineering {\em and} LLMs are depicted in Green ($|L \cap S|$). 
Given the rapid rise in overall publication volumes, we use a logarithmic scale for the vertical axis. 
Unsurprisingly, we see an overall rise in the number of CS publications. 

Also, given the recent upsurge in attention for LLMs, the exponential rise in the number of papers on LLMs is relatively unsurprising.

Perhaps more interesting is the rapid uptake of Software Engineering applications of LLMs, as revealed by the growth trend, pictured in green on this figure.
In order to  examine this trend in more detail, we plot the proportion of LLM publications ($L$) to all CS publications ($A$) in blue, as well as the proportions of LLM-based software engineering publications ($L\cap S$) to all LLM publications in orange in Figure~\ref{fig:trend-2}. 
As can be seen, the proportion of LLM papers on LLM-based Software Engineering has been rising dramatically since 2019.
Already, more than 10\% of all papers on LLMs are concerned with LLM-based Software Engineering.

As a result of this growth, we can expect many other surveys of LLM-Based SE. 
The rapid expansion of the literature makes it unlikely that further comprehensive SE-wide studies will fit the space constraints of a single paper, but we can expect many specific comprehensive surveys of sub-areas  of interest, and also Systematic Literature Reviews (SLRs) that tackle SE-wide crosscutting issues by asking specific research questions of the primary literature in the systematic review.
Already, such SLRs are appearing. 
For example, Hou et al.~\cite{hou2023large} provided an excellent recent SLR covering 229 research papers from 2017 to 2023 reporting SE tasks tackled, data collection and preprocessing techniques, and strategies for optimising LLM performance (such as prompt engineering).

The remainder of this paper is organised to follow the top-level software development activities and research domains as depicted in 
Figure~\ref{fig:structure}.

\section{Preliminaries}
\label{sec:preliminaries}

\subsection{Large Language Models}

A Large Language Model (LLM) refers to an Artificial Intelligence (AI) model that has been trained on large amounts of  data and is able to generate text in a human-like fashion.
Table~\ref{tab:term} provides a glossary of LLM  terminology to make the paper self-contained.

LLMs are typically based on deep learning techniques, such as transformers, and have the capability to generate useful language output.
As a result, they have been found capable of performing a wide range of language-related tasks, including text generation~\cite{goyal2023news}, answering questions~\cite{nakano2022webgpt}, translation~\cite{Brown2020}, summarization~\cite{xie2023survey}, and sentiment analysis~\cite{kheiri2023sentimentgpt}.

Rumelhart et al.~\cite{rumelhart1986learning} introduced the concept of Recurrent Neural Network, opening up the possibility of processing sequential data. Long Short Term Memory (LSTM) architecture, an extension of the RNN architecture introduced by Hochreiter and Schmidhuber~\cite{hochreiter1997long}, significantly improved their performance in many applications.

In 2017, Vaswani et al.~\cite{vaswani:attention} introduced the Transformer architecture, which captures word relationships with the self-attention mechanism.
The transformer architecture had a profound impact on language modelling and triggered an explosion of activity on LLMs.

In 2018, OpenAI released the Generative Pre-trained Transformer (GPT) model, followed by subsequent iterations (GPT-2, GPT-3, GPT-3.5, and GPT-4).
With GPT-3 and 3.5, many observers noticed  a significant step change in generative performance, thereby attracting a great deal of interest in GPT (and ChatGPT) in particular, and also in LLMs more generally.

LLMs achieve this performance, in part, due to the large corpora on which they are trained:
For example, GPT-3 is trained on 45TB of text data and has 175 billion parameters.
Meta launched open-sourced LLaMA in February 2023, which is trained on 1.4 trillion tokens with a variety of model sizes ranging from 7 billion to 65 billion parameters~\cite{touvron:llama}. 

\subsection{Categories of Large Language Models}
There are three categories of large language models: \\
\noindent 1) \textbf{Encoder-only model}: also known as an autoencoder, consists of an encoder network but does not have a separate decoder network. It takes an input sequence and maps it to a lower-dimensional representation. The purpose of an autoencoder is to learn an encoding of the input data. Examples of Encoder-only LLMs are BERT from Google, RoBERTa from Meta, and DeBERTa from Microsoft~\cite{yang_harnessing_2023}. \\
2) \textbf{Encoder-decoder model}: in addition to the encoder network, there is a decoder network that generates an output sequence by iteratively generating tokens or symbols based on the context vector and previously generated tokens. It can be adopted for tasks like machine translation or text generation.
Examples of Encoder-decoder LLMs are T5 from Google and BART from Meta~\cite{yang_harnessing_2023}.\\
3) \textbf{Decoder-only model}: Unlike the previous two types of LLMs, decoder-only LLMs do not have an encoder component to process the input data, but only a decoder component that directly generates an output sequence based on a given context or input.
Decoder-only models are often based on architectures such as autoregressive models, where the output is generated token-by-token. Each token generated by the decoder is conditioned on the previous tokens generated and the context.

Popular examples of decoder-only models are the GPT (Generative Pre-trained Transformer) series developed by OpenAI, LLaMA from Meta, Claude from Anthropic, and PaLM from Google~\cite{yang_harnessing_2023}. 

\begin{table*}[h!]
\caption{Existing Large Language Models for Code Generation}
\label{tab:existingLLMs}
\resizebox{0.99\textwidth}{!}{
\begin{tabular}{@{}llllllll@{}}
\toprule
\textbf{Name}          & \textbf{Release date} & \textbf{Produced by}                                         & \textbf{Parameters}       & \textbf{Open-sourced} & \textbf{Price}                                  & \textbf{Supported languages} & Type            \\ \midrule
\textbf{CodeBERT}      & February 2020         & Microsoft                                                    & 125M                      & YES                   & free                                            & 6                            & Encoder-decoder \\
\textbf{InCoder}       & April 2022            & Meta                                                         & 6.7B, 1.3B                & YES                   & free                                            & 30                           & Decoder-only    \\
\textbf{AlphaCode}     & February 2022         & DeepMind                                                     & 300M, 1B, 3B, 9B, and 41B & NO                    & free                                            & Python or C++                & Encoder-decoder \\
\textbf{CodeX}         & August 2021           & OpenAI                                                       & 12B                       & NO                    & free                                            & \textgreater{}11             & Decoder-only    \\
\textbf{Copilot}       & October 2021          & \begin{tabular}[c]{@{}l@{}}Github and  OpenAI\end{tabular} & 12B                       & NO                    & free for individual developers and organisations & \textgreater{}11             & Decoder-only    \\
\textbf{CodeT5}        & Nov 2021              & Salesforce Research                                          & 60M, 220M, 770M           & YES                   & free                                            & 6                            & Encoder-decoder \\
\textbf{CodeT5+}       & May 2023              & Salesforce Research                                          & 2B, 6B, 16B               & YES                   & free                                            & 9                            & Encoder-decoder \\
\textbf{PolyCoder}     & Oct 2022              & Carnegie Mellon University                                   & 160M, 400M, 2.7B          & YES                   & free                                            & \textgreater{}11             & Decoder-only    \\
\textbf{CodeWhisperer} & April 2023            & Amazon                                                       & unknown                   & NO                    & free for individual developers                  & 15                           & unknown         \\
\textbf{WizardCoder}   & June 2023             & Microsoft                                                    & 15B                       & YES                   & free                                            & unknown                      & Encoder-only    \\
\textbf{CodeGeeX}      & Sep 2022              & Tsinghua University et al.                                   & 13B                       & YES                   & free                                            & 23                           & Decoder-only    \\
\textbf{CodeGen}       & March 2022            & Salesforce Research                                          & 350M, 1B, 3B, 7B, 16B     & YES                   & free                                            & Python                       & Decoder-only    \\
\textbf{StarCoder}     & May 2023              & BigCode                                          & 15B                       & YES                   & free                                            & \textgreater{}80             & Encoder-only    \\
\textbf{phi-1}         & June 2023             & Microsoft                                                    & 1.3B                      & NOT YET               & free                                              & Python                       & Decoder-only    \\ 
\textbf{Code Llama}&August 2023&Meta&7B, 13B, 34B&YES&free&\textgreater{}7&Decoder-only\\
\bottomrule
\end{tabular}
}
\end{table*}

\subsection{Large Language Models for Software Engineering}
While LLMs have been widely applied to tasks involving natural languages, 
their application to software development tasks, involving programming languages, 
has also gained significant recent attention.

In 2021, OpenAI introduced CodeX, a fined-tuned descendant of GPT-3.
CodeX is used by GitHub's Copilot, which provides users of 
Visual Studio Code, Visual Studio, Neovim, and JetBrains with code completion. 
The new version of Copilot, GitHub Copilot X\footnote{GitHub Copilot X is under technical preview at the time we accessed it on July 17th 2023.}, is  based on GPT-4. 
In February 2023, GitHub reported that, on average, 46\%\footnote{In this paper, all percentages are reported with a precision of 2 significant digits.} of the developers' code was written by Copilot~\cite{copilot2023}.
For Java only, that number is 62\%. 
Thomas Dohmke, CEO of GitHub, said 
Copilot will write 80\% of code ``sooner than later'' in June 2023~\cite{soonerthanlater}.

In 2022, DeepMind introduced AlphaCode~\cite{li_competition-level_2022}, trained with 40B parameters on 
selected public GitHub repositories.
It achieved on average a ranking in the top 54\% in competitions with more than 5,000 participants in simulated evaluations.

The most recent GPT model, GPT-4, also performs code generation.
According to GPT-4's technical report~\cite{openai2023gpt4},
the zero-shot pass@1 accuracy is 67\% with GPT-4 on HumanEval, an open-source dataset from OpenAI consisting of 164 programming problems.

On a benchmark of 100 LeetCode problems,
GPT-4 has comparable performance with human developers~\cite{bubeck_sparks_2023}.
On the $24^{th.}$ August 2023, Meta released open-sourced Code Llama~\cite{codeLlama}, a state-of-the-art for publicly available LLMs on coding tasks.

Table~\ref{tab:existingLLMs} lists the representative LLMs that are designed for code generation/completion based on natural language descriptions.




\begin{table*}[h!]
\caption{Key Terminology Related to Large Language Models}
\label{tab:term}
\resizebox{0.99\textwidth}{!}{
\begin{tabular}{@{}lp{15cm}@{}}
\toprule
\textbf{Term}    &\textbf{Explanation }   \\ \midrule
Chain of Thoughts (CoT) & In the context of LLMs, chain of thought represents the logical flow of ideas and reasoning within the text generated by LLMs.             \\ \midrule
Encoder \& Decoder & Encoders are components of LLMs that map any given input of a specific type (such as image, audio, text) into a latent vector space. Decoders perform the reversal: they can take an input from a latent vector space, and (re)constructdata of the original type.\\ \midrule
Few-shot learning  & A machine learning technique that aims to train models to perform well on new tasks or classes with only a few new items of labelled training data. It is also known as in-context learning. With LLMs, few-shot examples are typically included in the prompt.            \\ \midrule
Fine-tuning        & A process by which a model, trained on a large dataset or a related task, is further trained on a smaller or more specific dataset to improve its performance on the target task or domain.                                 \\\midrule
Generative AI     & A category of artificial intelligence that focuses on generating or creating new content, such as images, text, music, and videos. \\ \midrule
Parameters & Parameters are the numerical values inside LLMs that are adjusted during training, and primarily include weights and biases. Weights dictate the strength of connections between neurons and serve as coefficients to the input values or activation thresholds for preceding neurons. Biases shift the weighted sum of inputs, before this sum is fed into the activation function. The number of parameters is often used as a measure of the size of an LLM. \\\midrule
Prompt & The input provided to the LLM to stimulate the generation of a response.                 \\\midrule
Prompt engineering & The process of designing and optimising prompts to achieve desired outcomes.                                            \\\midrule
ReAct & The ReAct (Reasoning and Acting) prompting framework allows LLMs to generate reasoning traces as well as actions specific to the given task. Once an LLM generates an action, it can be carried out externally, and the observation of the output of the action can be included in the next prompt, providing information to the LLM. This enables LLMs to use external tools.\\\midrule
Temperature & A parameter that affects the randomness of the generated content. A higher value (e.g., 1.0) yields more diverse and creative content, while a lower value (e.g., 0.2) yields more deterministic content.\\\midrule
Token & A token is the atomic unit with which an LLM represents its input and output. Tokens are enumerations, and can represent words, characters, subwords or other segments of text and/or code.\\\midrule
Top-N, Pass@N & For many applications, LLMs will typically generate a number of candidate solutions in a ranking. Top-N metrics count the number of problems correctly solved by an LLM with an answer among its Top N candidates. Similarly, Pass@N counts the number of programming questions for which a candidate program within the Top N rank has passed the test case.\\ \midrule
Zero-shot learning & A machine learning technique that enables models to generalize and make predictions on classes or tasks that were not seen during the training phase. There is no labelled data available for these new classes. \\  

\bottomrule
\end{tabular}
}
\end{table*}

\section{Requirements Engineering and Design}
\label{sec:design}
Requirements engineering is an important discipline in software engineering.
It forms the fundamental link between the technical attributes of the system software engineers build, and the purpose for which the systems are built.
There is a mature literature, and a large research community concerned specifically with problems associated with requirements engineering problems \cite{neill:requirements}.

There has also been previous work on artificial intelligence approaches to support requirements engineering, notably in the form of computational search for requirements engineering \cite{yyetal:refsq08}.
However, hitherto, the discipline of requirements engineering has received less attention from the emerging literature on LLM-based software engineering.

Zhang et al.~\cite{zhang_preliminary_2023} conducted a preliminary evaluation of ChatGPT's zero-shot requirement retrieval performance on two requirements analysis tasks over four data sets. 
Although these results are only preliminary, they provide optimism that LLMs can be used as a support for efficient and effective requirements engineering.
Luo et al.~\cite{luo2022prcbert}
conducted prompt engineering with BERT for automatic requirement classification. 
Luitel et al.~\cite{luitel2023improving} focused on requirements completeness and 
used
BERT to generate
predictions for filling masked slots in requirements.

\subsection{Open Problems in LLMs for Requirement Engineering}
Unlike other software development activities, we did not find much work on LLM-based requirements engineering or on LLM-based design. 
Indeed, there was even evidence that practising engineers are reluctant to rely on LLMs for higher-level design goals~\cite{Liang:large-scale}.
There is thus a great opportunity to expand on this open field of research.

The majority of LLM applications are focused on tasks such as code generation, testing, and repair.
These tasks benefit from LLM's  capability to generate code.
Nevertheless, LLMs also have significant potential to support 
requirements engineering activities, thanks to their natural language processing capabilities.

For example, traceability is
a long-standing, cross-cutting 
concern in software engineering. 
In particular, identifying traceability links between requirements 
and other engineering artefacts, such as code and tests, are especially challenging because requirements are often written in natural 
language; a natural fit for LLMs.

\section{Code Generation and Completion}

\spec{Of all the application areas, code completion is one that is most thoroughly explored already.
We will provide a survey of this existing work.
The focus will be on the remaining challenges for code completion but also considering challenges for generating new features, and the degree of engineering guidance required to ensure the code is as expected by the software engineer. There is some overlap here, with repair refactoring and performance improvement, all of which involve generating code, but deserve sections in their own right.}

Of all the Software Engineering application areas for LLMs, 
code completion is the area that has been most thoroughly explored hitherto.
Even prior to the advent of LLMs, it was suggested that learning from existing code repositories is the key to successful and intelligent code completion~\cite{bruch:completion}: pre-trained LLMs deliver on these early aspirations for code completion. 
While hallucination has been pointed out as the weakness of LLMs more generally, the specific task of code completion sidesteps hallucination problems by acting as a recommender system to the developer.
The developer thus bears the responsibility to weed out any hallucinated LLM output before it leaks into the code base.

Of course, a high degree of hallucination would have made code completion recommender systems ineffective.
The widespread and rapid adoption, and the positive results already reported for code completion,
provide early indications that this has not happened. For example, Murali et al.~\cite{murali:codecompose} reported the experience of deploying CodeCompose, a code completion tool based on the Incoder LLM~\cite{fried:incoder}, at Meta. During 15 days, 4.5 million code completion suggestions were made by CodeCompose, and the acceptance rate from developers was 22\%. 
The qualitative feedback was highly positive, with 92\%  positive. Similarly, Peng et al.~\cite{peng:impact} reported that programmers could complete a non-trivial task (implementing an HTTP server in JavaScript) 56\% faster when paired with GitHub Copilot, compared to the control group that did not receive any such support.

Many software engineers already appear to have decided that benefits outweigh any necessary human filtration effort,
with enthusiastic levels and rates of adoption already being reported. Once LLM-based code completion is fully adopted, there are expectations that programmers will spend more time reviewing rather than writing code~\cite{bird2022taking}.

\subsection{Code Generation Models}
Automated code generation has a long history, tracing its origins back to early visions of automated program synthesis \cite{manna:toward}, which have continued to develop and have generated impressive results \cite{gulwani:program-synthesis}.

From the pioneering work of Hindle et al. on the naturalness of software
 \cite{hindle:naturalness}, we know that programmers write code (and languages enforce code writing styles), that make code highly predictable.
Furthermore, Barr et al. \cite{ebetal:psh-fse14} found that  43\% of commits to a large repository of Java projects could be reconstituted from existing code.
They called this  `The Plastic Surgery Hypothesis' because of the way automated repair proceeds by scavenging for existing code to patch up issues elsewhere  \cite{legoues:cacm-survey}.

Their empirical study  provided  evidence for the efficacy of this scavenging approach, but also underlined the repetitive and predictable nature of software.
In a larger repository (sourceforge), Gabel and Su \cite{gabel:uniqueness} 
found that a programmer would have to write
more than six lines of code in order to create a novel code
fragment.

These research findings on code naturalness, reusability and predictability, make it unsurprising that LLMs have been able to 
exploit that same predictable reusability to produce effective recommendations for code generation.
These observations have underpinned the growth of generate-and-test approaches to repair and genetic improvement \cite{legoues:cacm-survey,Petke:gisurvey}.
The generate-and-test approach offers greater code transformation freedom (compared to more traditional  correct-by-construction approaches~\cite{darlington:transformation}), precisely because the generated code may not preserve strict, mathematically-defined  (and not always appropriate, nor useful) interpretations of correctness.

This freedom to explore a wider space of ``semantic near neighbours'' allows Genetic Improvement to find dramatic optimisations (see Section~\ref{sec:perf}). 
The Genetic Improvement approach, nomenclature, and evaluation methodologies also provide a scientific framework within which to understand and evaluate LLM-based code generation.
Both technologies 
share the `generate-and-test' approach to program transformation and code generation,
potentially making much of the existing work on genetic improvement directly applicable
to LLM-based code generation.

In 2021, Chen et al.~\cite{chen_evaluating_2021-short} introduced CodeX, a GPT language model fine-tuned on publicly available code from GitHub, and evaluated its Python code-writing capabilities. 
They released a new evaluation set called `HumanEval' to measure functional correctness for synthesizing programs from docstrings, and found that CodeX outperformed GPT-3 and GPT-J when tackling these problems. 
Since then there has been an explosion in research on LLM-based code generation and the HumanEval dataset has been used in many subsequent studies.

In 2022, Li et al.~\cite{li_competition-level_2022} introduced AlphaCode, a system for code generation that  creates novel solutions to competitive programming problems. 
They found that three key components were critical to achieving reliable performance: 

\begin{enumerate}
    \item An extensive  programming dataset for training and evaluation. 
    \item Large and efficient-to-sample transformer-based architectures.
    \item Large-scale model sampling to explore the search space, followed by behaviour-based filtering.
\end{enumerate} 

In simulated evaluations on programming competitions on the Codeforces platform, AlphaCode achieved, on average, a ranking of the top 54\% in competitions with more than 5,000 participants.

Several papers also introduced code synthesis LLMs~\cite{nijkamp:codegen,nijkamp:codegen2,xu_systematic_2022,jiang_discovering_2022}, based on large data sets with little pre-filtering of the training data. 
However, in 2023, Gunasekar et al.~\cite{gunasekar2023textbooks} reported that, by training with only a textbook-quality code corpus, LLMs with lower parameter counts could achieve performance comparable to much larger models. 

They classified an existing Python code corpus with the GPT-4 model, by prompting it to determine the educational value of the given code for a student who wants to learn programming. Second, they used GPT-3.5 to create synthetic textbooks about Python.
Specific code generation use cases have also been tackled, such as numerical algorithm code generation ~\cite{kashefi_chatgpt_2023}, and generation of code from behavioural descriptions \cite{chemnitz:code}.
More examples of the existing LLMs for code generation and the code generation leaderboard can be found in Table~\ref{tab:existingLLMs} and Figure~\ref{fig:pass1}.

\subsection{Prompt Engineering for Improved 
Code Generation}
Prompt engineering has been extensively used as a way to improve code generation.
For example, Li et al.~\cite{li_towards_2023} reported
 pass@1 improvements of between approximately 50\% and 80\% on
CodeX, CodeGeeX, CodeGen, and InCoder on several benchmarks 
(MBPP for Python, MBJP  for Java, and MBJSP  for JavaScript).
D\"{o}derlein et al.~\cite{doderlein_piloting_2023} 
reported the prompt-engineered improvement of Copilot and CodeX success rates from approximately 1/4 to 3/4 on HumanEval and LeetCode.
He and Vechev~\cite{he_controlling_2023}
used prompt engineering to improve the security of LLM-generated code, reporting an improvement in security from 59\% (of cases considered)  to 92\%. 
White et al.~\cite{white_chatgpt_2023} provided a catalogue of prompt engineering design patterns for various software engineering tasks, including code generation.
Denny et al.~\cite{denny2023conversing} argued that prompt engineering is a useful learning activity that fosters software engineering students' computational thinking.


Other authors have considered ways to decompose prompt engineering into iterative and multiphase conversations with the LLM, moving it closer to Chain of Thought reasoning.
For example, Li et al.~\cite{li:skcoder,li:enabling} reported an 18\% increase in ChatGPT Pass@1 using a two-stage sketch-based approach, SkCoder, in which the LLM first creates a sketch and then subsequently implements these sketches.
Jiang et. al.~\cite{jiang2023selfevolve} and Zhang et al.~\cite{zhang2023selfedit}  also sought to deploy Chain-of-Thought-style reasoning by prompting LLMs to reflect and self-edit.

Existing software engineering analysis techniques can also provide additional information for fine-tuning and prompt engineering. 
For example, Ahmed et al. \cite{ahmed_improving_2023} show how simple static analysis can be used in the prompt to improve the performance of code generation with few-shot learning.

Shin et al.~\cite{shin2023prompt} compared 
prompt engineering and fine tuning with GPT-4 for code generation tasks, demonstrating that fine-tuning works better than prompt engineering.

\subsection{Hybrids of LLMs and other Techniques}
Throughout our survey of the literature, we found strong evidence that some of the most promising results can be achieved by hybridising; combining LLMs with other existing software engineering techniques.
This section surveys work on hybrid LLMs for code generation.

Several authors have developed hybrids of LLMs combined with planning and search.
For example, 
Zhang et al.~\cite{zhang2023planning, jiang_self-planning_2023} reported improvements over baselines of between approximately 11\% and 27\%, while
Zhang et al.~\cite{zhang:toolcoder} hybridized code generation with API search techniques. 

Hybrid approaches have also 
used existing software engineering and/or AI techniques 
to select the best candidate from an LLM's top-n outputs. 
For example, Chen et al.~\cite{chen_codet_2023} use test generation to choose candidates and reported improvement of approximately 20\% 
on five pre-trained LLMs; 
Inala et al.~\cite{inala:faultaware} use a neural network-based ranker
to predict code correctness and potential faults.
Jain et al.~\cite{jain_jigsaw_2022} proposed Jigsaw, which 
post-processes the generated code based on
program analysis and synthesis techniques.

Dong et al.\cite{dong_self-collaboration_2023} treated LLMs as agents, letting
multiple LLMs play distinct roles in addressing code generation tasks collaboratively and interactively.
They reported improvements of approximately 30\%-47\%.

\subsection{Scientific Evaluation of LLM-based Code Generation}
There is a pressing need for more thorough scientific evaluation.
Many authors have anecdotally reported on cases where LLMs failed to generate correct, secure, and reliable code.
Poldrack et al.~\cite{poldrack_ai-assisted_2023}  also highlight the need for substantial human validation.
In this section, we survey the literature on the empirical evaluation of LLM-based code generation in terms of correctness,  
robustness,
explainability,
determinism,
and security.


\subsubsection{Correctness Evaluation}
The GPT-4 Technical Report~\cite{openai2023gpt4}
evaluated the correctness of GPT-4's code generation on
the HumanEval dataset, reporting a zero-shot accuracy of 67\%, a modest improvement on
the (earlier ChatGPT) results reported by Yetistiren et al. \cite{yetistiren_evaluating_2023}.

Borji~\cite{borji_categorical_2023} presented a rigorous, categorised and systematic analysis of LLM code generation failures for  ChatGPT. 
Eleven categories of failures, including reasoning, factual errors, mathematics, coding, and bias, are presented and discussed in their work.

\begin{figure*}[tb]
\centerline{\includegraphics[width=0.90\linewidth]{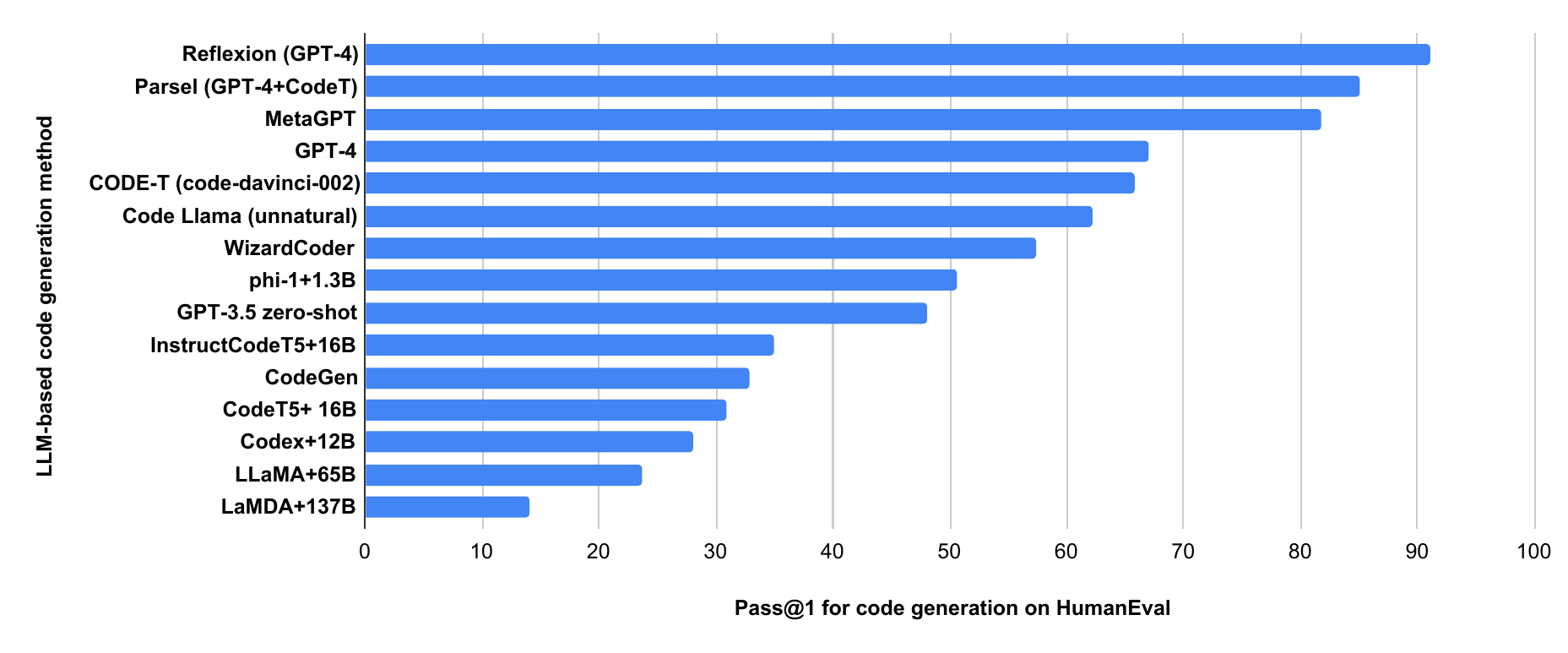}}
\vspace{-5mm}
\caption[]{Code generation leaderboard for the HumanEval benchmark. These methods are either based on LLMs or LLMs themselves.\label{fig:pass1}}
\end{figure*}

Figure~\ref{fig:pass1}
shows the leaderboard of code generation correctness in terms of the pass@1 (i.e., the test pass rate for the top-1 code candidate) on the HumanEval dataset according to Papers With Code, a platform that highlights trending AI research and the code behind the method and models.\footnote{The actual leaderboard can be found at \url{https://paperswithcode.com/sota/code-generation-on-humaneval/}; results in Figure~\ref{fig:pass1} accessed on 24th August 2023.}
The LLM models behind each method are shown in brackets. 
At the time of writing, the best code generation model, Reflexion~\cite{shinn_reflexion_2023}, can generate correct code for over 90\% of the generation tasks.  
However, these numbers and the relative rankings of different language models are inherently subject to change in such a rapidly developing field. 
For example, the figure given for correct code on HumanEval in the original GPT-4 Report~\cite{openai2023gpt4} was only 67\%, so the updated figure of 80\% 
(at the time of writing, which is five months later) retrieved from the Papers-With-Code website presumably represents the evolution of GPT4 since then.


Despite the promising results in the literature on code generation and completion,
Din et al.~\cite{dinh2023large} reported that the performance of code completion dropped by more than 50\% on HumanEval when the context contains bugs.

\subsubsection{Robustness Evaluation}
LLM code generation robustness is the degree to which similar prompts elicit semantically and syntactically similar code generation.
Treude~\cite{treude_navigating_2023} introduced GPTCOMPARE, a prototype tool for visually highlighting similarities and differences between LLM code outputs.
Yan et al.~\cite{yan2023coco} introduced COCO to test the robustness and consistency of LLM-based code generation systems. 

\subsubsection{Explainability Evaluation}
\label{sec:explanability}
One considerable advantage of LLMs, over previous machine learning techniques, is the way in which the code generation artefacts are accompanied by explanations.
Such explanations have the potential to increase adoption, by providing additional confidence and faster understanding.
More work is needed to evaluate and optimise explanations that accompany generated code and other software engineering artefacts.

Initial evaluation by MacNeil et al.~\cite{macneil_experiences_2023} 
on their interactive Web development e-book, suggested that a majority of students perceived LLM-generated code explanations to be helpful.
Noever and Williams~\cite{noever_chatbots_nodate} also showed the potential for explanations to help human engineers, particularly where code is obfuscated or lacks sufficient existing documentation.
In this way, the ability to produce insight and explanation may go beyond simply justifying the code generated by the LLM itself, but may become a valuable source of education and documentation (See Section~\ref{sec:education}).

Sun et al.~\cite{sun_investigating_2022} focus on users’ explainability needs for generative AI in three software engineering use cases: code generation based on natural language description (with Copilot), translation between different programming languages (with Transcoder), and code autocompletion (with Copilot). 
Their investigation was conducted as 9 workshops with 43 software engineers and identified 11 categories of explainability needs in the context of Generative AI (GenAI) for code. It also proposed 4 types of  features for generative AI: AI documentation, model uncertainty, attention distribution, and social transparency (i.e., making visible the socio-organizational factors that govern the use of AI).

Mohammadkhani et al.~\cite{mohammadkhani2022explainable}used the attention mechanism to study 
CodeBERT and GraphCodeBERT on tasks including code documentation generation,
code refinement, and code translation.

\subsubsection{Determinism Evaluation}
\label{sec:determinism}
LLMs are nondeterministic.
Ouyang et al.~\cite{ouyang2023llm} empirically studied the non-determinism of ChatGPT in code generation, founding that 
over 60\% of tasks had zero equal test output across different requests.
Nevertheless, their study of the literature in
LLM-based code generation 
demonstrate that only 21.1\% of these
papers consider the non-determinism threat in their experiments.

\subsubsection{Security Evaluation}

Hajipour et al. \cite{hajipour_systematically_2023} proposed a few-shot prompting approach to detecting security vulnerabilities, reporting that their approach automatically finds thousands of security vulnerabilities in several models.
Khoury et al.~\cite{khoury_how_2023} found that the code generated by ChatGPT often fell way below even minimal standards of secure coding. 
Risse and  B\"{o}me~\cite{risse:limits} reported results that indicated vulnerability detection accuracy may be over-reported, due to the model overfitting to unrelated training set features
.

In addition, Yetistiren et al. \cite{yetistiren_evaluating_2023} 
presented a comprehensive evaluation of the performance of 
Copilot, 
CodeWhisperer, 
and ChatGPT, covering different aspects including 
code validity, 
code correctness, 
code security, 
code reliability, 
and 
Their results show a wide degree of divergence in performance, 
motivating the need for further research and investigation. 
For example, they found 65\%, 46\%, and 31\% of the programs generated by
ChatGPT, Copilot, and CodeWhisperer (respectively) were correct.

\subsubsection{Benchmarks}
As with other scientific evaluations, software engineering evaluation relies on publicly available and representative benchmark suites.
A number of these have already emerged and can support software engineering evaluation of LLM-based applications.
The Papers-With-Code platform\footnote{\url{https://paperswithcode.com/task/code-generation}} provides a summary of 15 benchmarks for evaluating code generation.

Evaluations have often relied on small programming problems from programming courses \cite{savelka_can_2023}, 
synthetically generated problem sets
\cite{liu_comprehensive_2023},
and online judging platforms such as Leetcode \cite{bubeck_sparks_2023,zhang2023selfedit,nguyen:empirical-copilot}. 
Although results reported naturally vary by LLM in training sets, the overall conclusions of these evaluations indicate success rates of between 20\% and 80\%.

Nevertheless, existing code generation benchmarks tend to rely on test suites to automatically judge code correctness, which can be inadequate, leading to false judgements~\cite{liu:judges}.
This highlights the need for more work on evaluation benchmarks that are specifically tailored to LLM-based code generation evaluation.
Liu et al.~\cite{liu_is_2023} draw attention to the problem, showing how existing test suites can lead to high degrees of false positive conclusions (also a serious problem for online judge platforms~\cite{liu:judges}).
To alleviate this problem,
they propose EvalPlus – a code synthesis benchmarking framework that automatically generates test inputs and rigorously evaluates the functional correctness of LLM-generated code.
Their evaluation of 14 popular LLMs (including GPT-4 and ChatGPT) demonstrated that with the newly generated tests for HumanEval, the assessment of pass@k drops by up to 15\%, averaged over problems considered.

Jimenez et al.~\cite{jimenez2023swe} 
introduced SWE-bench with the aim of evaluating LLMs on code generation problems in a realistic software engineering setting.
SWE-bench contains 2,294 software engineering problems, drawn from real GitHub issues.
The results suggest that Claude 2 and GPT-4 solve only 4.8\% and 1.7\% of the coding tasks, respectively.


\subsection{Open Problems in Code Generation and Completion} 

Assessing the generated code remains a critical problem for LLM-based code generation and completion: while much work already started applying existing software testing knowledge to this problem, we expect closer integration of automated testing techniques with code generation and completion techniques. 

Fortunately, there is a large body of existing work on automated test data generation \cite{anandetal:orchestrated,cadar:three-decades,mh:icst15-keynote}, much of which will have an important role to play in ensuring the correctness of the engineering artefacts generated by LLMs.
A recurring theme of the challenges covered in this paper, is that code execution provides precisely the `ground truth' needed to filter hallucinated responses. 
It can also provide guidance as part of interactive reasoning/action (`ReAct') dialogue \cite{yao:react}, both with and within LLMs.

Automated test data generation allows the software engineer to target the exploration of the most relevant regions of this run-time ground truth.
This test-based targeting can help filter, fine-tune and to optimise prompts, thereby minimising problems posed by hallucination.
LLMs also have considerable potential for automating the process of constructing effective and efficient software test suites.

Another important problem is how to efficiently fine-tune pre-trained LLMs so that they perform better for a specific programming language, codebase, or domain: this is especially important because training an LLM from scratch requires significant computational resources. 

For example, transfer learning has been proposed as a way to improve code completion performance when the volume of training examples for a specific programming language is inadequate~\cite{zhou:improving}.

The current focus of research is on the code produced by LLMs.
However, the explanations produced by LLMs may turn out to be at least as important.
One could imagine many scenarios in which an engineer would prefer to accept a (possibly) suboptimal software engineering artefact that comes with a compelling explanation, over a potentially more performant solution with a less compelling explanation.
After all, engineers regularly make the same judgement call for human-designed engineering artefacts, so  why would we expect it to be any different for those produced by machines?
As with prompt engineering, which focuses on optimising the input to the LLM, explanation engineering is also likely to become an area of study in its own right.

\section{Software Testing}
\spec{software test generation is already well explored using non-LLM. However, LLMs have the potential to generalise from existing tests (for example to increase coverage and to tackle corner cases that may be missed). When generating new tests we have the challenge of determining whether the test assertion is itself correct. The section will tackle known problems for software test generation such as the Oracle problem, and coverage challenges. The section will also explore the possibility of including dynamic information from execution for prompt engineering. The section will also briefly review existing test generation techniques, and how these can be used as a foundation for all of the remaining sections in the paper, as explained in the abstract. The section also surveys existing work on AI-based test generation.}

Software testing is a well-established research discipline, the origins of which 
can be traced back to Turing's pioneering work in the late 1940s \cite{turing:checking}.
Much of the focus of this research has been on the automated generation of test suites, able to achieve high fault revelation potential at low computational cost \cite{anandetal:orchestrated,cadar:three-decades,mh:icst15-keynote}.
This provides us with, not only  techniques able to weed out incorrect LLM-generated code, but also a mature baseline against which to compare novel LLM-based and hybrid techniques for test suite generation.

There is already a sufficiently large body of work to warrant a survey specifically on LLM-based Software Testing: 
Wang et al. \cite{wang:llm-test-survey} presented a survey of papers primarily on testing, but also including debugging and repair. 
They reported on 52 papers (33 published since 2022) of which approximately one-third concerned test-based LLM fine-tuning, while the remainder relied upon prompt engineering.

\subsection{Generating New Tests Using LLMs}
In this section, we review existing work on LLMs for test data generation, before highlighting open problems and challenges for the development of this emerging field.
The tests generated may not be executable because the LLM is not guaranteed to generate compilable code.
Nie et al. \cite{nie:learning} report 29\% of tests generated using TeCo are executable,
while
Yuan et al.~\cite{yuan_no_2023}  found that approximately one-quarter of the tests generated by 
ChatGPT were executable, rising to one-third with suitable prompt engineering.

Of those tests that do compile, several authors have reported on the code coverage achieved. 
For example,
Barei{\ss} et al.~\cite{bareiss2022code}
reported an increase from the 10\% achieved using
Randoop \cite{pacheco:randoop} to 14\% with CodeX.
Hashtroudi et al.~\cite{hashtroudi2023automated} reported a 50\% increase in line coverage for the tests they generated by fune-tuning CodeT5.
Siddiq et al.~\cite{siddiq_exploring_2023}
reported 80\% coverage on
the HumanEval dataset using CodeX, 
but also found that neither the studied LLMs
could achieve more than 2\% coverage on the EvoSuite SF110 dataset.

Hybrid approaches that combine existing test generation and evaluation techniques, such as fuzz-based testing and search-based testing, with LLMs have already demonstrated promising results. 
For example,
Lemieux et al.~\cite{lemieux_codamosa_nodate} introduced CODAMOSA, an algorithm that combines Search-Based Software Testing (SBST) \cite{mh:icst15-keynote} and CodeX to generate high-coverage test cases for programs under test. 
When SBST's coverage improvements stall, CODAMOSA asks CodeX to provide example test cases for under-covered functions.
This helps SBST redirect its search to more useful areas of the search space. In an evaluation of 486 benchmarks, CODAMOSA achieved significantly higher coverage compared to SBST and LLM-only baselines. 
Hu et al.~\cite{hu2023augmenting} introduced ChatFuzz, which augments the widely studied fuzzer, AFL, with ChatGPT, in order to get more format-conforming mutants. In an evaluation of 12 target programs chosen from three benchmarks, ChatFuzz achieved higher branch coverage than AFL by 13\%.
Dakhel et al.~\cite{moradi2023effective} used mutation testing to help LLMs to generate tests. 
In particular, they 
augmented prompts for Codex and Llama-2-chat
 with surviving mutants.
They report that their approach detects 28\% more human-written faults.
Xia et al.~\cite{xia2023universal} recently demonstrate that LLMs can serve as a universal fuzzer for systems across different application domains and programming languages, including C/C++ compilers, JDK, SMT solvers, and even quantum computing systems.

Deng et al.~\cite{deng_large_2023} propose TitanFuzz, which uses LLMs (i.e., Codex) to generate valid input DL programs to test DL libraries. 
The results on PyTorch and TensorFlow reveal that TitanFuzz can achieve
30\%/51\% higher code coverage than state-of-the-art fuzzers.
Later on, they further introduced FuzzGPT~\cite{deng_large_2023new}, which synthesizes unusual programs for fuzzing DL libraries. 
Their results indicated that CodeX and CodeGen could outperform TitanFuzz on PyTorch and TensorFlow when re-targeted for fuzz-based testing.

Li et al. \cite{li_finding_2023} used a hybrid of differential testing and ChatGPT
to elevate the latter's ability to generate failure-inducing test cases of buggy programs.
They report a test effectiveness improvement from 29\% to 78\%.



A promising area for LLM-based test generation is GUI testing, because the manipulation of the application state via GUI often requires a semantic understanding of both the user interface as well as the 
application domain. 
Sun et al.~\cite{sun2023automatic} 
described user interface via text, and asked ChatGPT which action it would like to perform next based on the text, then convert the answer into actual GUI interaction. This resulted in 32\% higher activity coverage compared 
to the state-of-the-art.

One particularly important problem that is challenging for classical techniques is the construction of test cases from user reports.
The user reports are written in natural language. 
This has presented considerable challenges for existing techniques, but is ideally suited to LLMs.
Kang et al.~\cite{kang2022large} introduced Libro, a  few-shot learning failure reproduction technique that automatically generates tests from
general bug reports, based on CodeX.
Libro successfully reproduced  
approximately one third of the failures.

Feng and Chen~\cite{Feng:prompting} demonstrated a replicability rate  of 80\% on  bug reports with natural-language-defined steps to reproduce, using an LLM out of the box (ChatGPT) with Chain 
of Thought prompt engineering alone.

Several authors have considered prompt engineering to improve the results of test generation
\cite{yuan2023manual,schafer_adaptive_2023}. 
For example,
Schafer et al.~\cite{schafer_adaptive_2023} proposed TESTPILOT, which re-prompts with failing tests and associated error messages, achieving reported 
average statement coverage of 68\%.
Xie et al.~\cite{xie_chatunitest_2023}
create prompts for test generation by 
 parsing the project and creating an adaptive focal context that includes the focal method and its dependencies.
They further used rule-based repair to ﬁx syntactic and simple compile errors in the tests.


Although the outcomes of LLM-based testing may be uncertain, 
researchers have explored cross reference or majority of votes~\cite{zsetal:automatic,zsetal:improving} methods to estimate the confidence of LLMs, based on the notion of `self-consistency'~\cite{wang2023selfconsistency}. 
For example, the Libro introduced by Kang et al.~\cite{kang2022large} uses CodeX to generate tests from bug reports that can reproduce failures. 
If multiple tests show similar failure behavior, Libro estimates that LLM is ``confident'' in its predictions.
Furthermore, where there is partial oracle information,  this can also be used to augment confidence estimates.
Such partial oracle information is often available when the goal of the overall processes to improve on existing code. 
For example, when improving the efficiency of an existing test, automated partial oracle information can be gathered from observing whether the test behaves similarly to the original (passing and failing in the same situations), and is also faster to execute.

\subsection{Test Adequacy Evaluation}
Test effectiveness is typically measured in terms of `adequacy criteria' \cite{ostrand:dataflow,bertolino:testing}.
Since testing cannot exhaustively explore every possibility, adequacy criteria provide a form of lower bound on the effectiveness achieved by a suite of tests.
Mutation testing is a widely-studied technique for assessing the adequacy of software test suites \cite{yjmh:analysis,mpetal:mutation-advances}, in which synthetic faults (called `mutants'), are deliberately injected in order to assess test adequacy. 
Mutation testing has been shown to provide more stringent adequacy criteria than other structural coverage-based criteria such as statement and branch coverage \cite{mike:icse17}.

One of the challenging open problems for mutation testing is to generate mutants that faithfully model important classes of real-world faults.
Khanfir et al.~\cite{khanfir2023efficient} used CodeBert to generate developer-like mutants and found that their approach has better fault revelation ability than PiTest.
Garg et al.~\cite{garg_vulnerability_2023}  applied CodeBERT to generate mutants that faithfully capture vulnerabilities.
They evaluation
found that 17\% of the mutants fail the tests that are failed by 89\% of the respective vulnerabilities.
Brownlee~\cite{Brownlee:2023:SSBSE} used GPT-3.5 to generate mutants for genetic improvemnt and observed that randomly sampled LLM-based edits compiled and passed unit tests more
often compared to standard GI edits. 

\subsection{Test Minimisation}
Test minimisation improves the efficiency of software testing by removing redundant test cases.
Pan et al.~\cite{pan2023ltm} applied
CodeBERT, GraphCodeBERT, and UniXcoder to extract
embeddings of test code to conduct test minimisation. 
Their approach achieves a 0.84 fault detection rate and runs much faster (26.73 minutes on average) than the baseline.

\subsection{Test Output Prediction}
Liu et al.~\cite{liu_code_2023} proposed CodeExecutor, a pre-trained Transformer model, to 
predict the program’s whole execution trace. The purpose is to imitate the real-world arbitrary program execution behaviour.
Their evaluation compares CodeExecutor with CodeX, and shows that CodeExecutor significantly outperforms Codex in execution trace prediction (e.g., 76\% vs. 13\% output accuracy for the Tutorial dataset).

\subsection{Test Flakiness} 
A test is flaky if it can pass on some occasions and fail on others without any apparent (tester-controllable) change in the execution context.
Test flakiness is one of the most pressing and impactful problems that inhibit test effectiveness in industry \cite{mhpoh:scam18-keynote}.
LLMs have been used to predict flakiness with high accuracy (with 73\% F1 score \cite{akli:flakycat,fatima2022flakify} and 97\% accuracy \cite{fatima:blackbox} reported). 

\subsection{Open problems in LLMs for Software Testing}
There are many open problems in LLM-based software test data generation, most of 
which lie well within the grasp of existing software testing techniques. 
We can thus expect an exciting explosion in LLM-based software test generation in the coming years.
This section outlines some directions for this research agenda.

\subsubsection{Prompt Engineering}
There are many aspects of a good software test that could be favoured by suitable prompt engineering. 
For example, we need to understand how to engineer prompts that 

\begin{itemize}
\item Predict and reduce generated test flakiness;
\item Reveal likely faults, for example via training on historic fault data;
\item Optimise the balance between mocking and integration testing;
\item Make realistic data builders, mock objects, parameters and inputs;
\item Predict tests that are most likely to elicit tests that cover corner cases;
\item Tailor test generation to focus behaviour that is prevalent in production.
\end{itemize}

\subsubsection{Augmenting Existing Tests}
Work on LLM-based test generation has focused on the automated generation of novel test suites. However, given the  array of existing test generation techniques, there remains an important (and comparatively less well-studied) open problem of augmentation and regeneration based on existing test suites~\cite{santelices:augmentation,symh:regeneration}.

Test augmentation and regeneration can exploit few-shot learning and/or can fine-tune (on an existing suite of test data and historical faults), to generate augmented test suites.

More work is needed on LLMs for generating additional test assertions that capture corner cases, historical faults, and likely programmer errors, drawing on the training data available. Hybridization 
between LLMs and existing automated test generation techniques is also a productive theme~\cite{lemieux_codamosa_nodate}.

\subsubsection{Test Correctness}
Traditional software test generation has suffered from the Oracle Problem~\cite{ebetal:oracle}, i.e., they are inhibited by the lack of an automated oracle that determines whether a test outcome is correct. Two cases pertain to AI-generated tests:

\begin{enumerate}
    \item {\bf The generated test passes on the current release}: We might assume that the functionality is correctly tested and that the generated test thus acts as a regression test, against which future changes can be checked.
    \item {\bf The generated test fails on the current release}: We need to know whether the assertion is wrong or whether the generated test has found a bug. 
\end{enumerate}

Both cases can have pernicious consequences when they are not imbued with self-regulation. A test case that passes may merely reflect coincidental correctness \cite{abou:coincidental,kaetal:analysis}.
Worse, it might be the case that the code is, in fact, {\em incorrect} (and that the test is equally incorrect yet captures, and thereby enforces, the incorrect behaviour). 
In such cases, the generation of the 
test will tend to {\em inhibit} fault remediation, by failing on future fixes. This problem also affects LLM-generated test cases, and may be more pernicious in cases where such tests hallucinate 
oracle properties, baking into  the generated tests these incorrect oracle assertions.

On the other hand, when a generated test case fails, this may indicate a bug. This bug revelation would denote a `win' for LLM-based testing. However, should it turn out that the ratio of false 
positives to true positives are high, then the cost (e.g., in human assessment) may make the technique impractical, even when it does reveal true positive bugs~\cite{mhpoh:scam18-keynote}.
More work is needed on self-assessment of confidence, self-checking for correctness, consistency, and robustness of generated tests.
We need to develop techniques for automatically assessing, augmenting and 
filtering raw outcomes from execution of LLM-based tests, before presenting the `test signal' to the developer.

The interaction between LLM hallucination and test correctness is an important topic in its own right. 
Since LLM-based code generation is generally driven by what is most likely, rather than what 
is most correct, hallucination poses threats to any questions of correctness. 
However, interestingly, Feldt et al.~\cite{feldt:autonomous} reported a case of hallucination being {\em helpful} for software 
testing, because it may reveal discrepancies between the actual program semantics and the programmer's perception of the semantics. 
They suggested a form of conversational testing agents 
(i.e., any generated 
tests are filtered by the programmer via the conversation) to harness this capability without posing any threats to overall test correctness.

More work is also required on the scientific foundations on which evaluations of LLM-based software testing rest.
More care and attention are clearly needed to heed the `best practice' advice for the scientific analysis and reporting from
previous work on foundations of Empirical and Search Based Software Engineering \cite{easterbrook:selecting,arcuri:practical,mhetal:sbse-tutorial}.

\subsubsection{Mutation Testing}

More work is needed to explore the adequacy achievable with LLM-based test generation, and also to use LLM-based techniques to support and enhance test adequacy investigation and assessment.
LLMs can be fine-tuned on a fault model, and thereby used to suggest mutants that are highly coupled to real faults, and can thus be used to assess test adequacy.

\section{Maintenance, Evolution and Deployment}
Software maintenance and evolution have been important topics of study for many decades.
They are concerned with existing code bases from which we seek understanding and business logic extraction, and for which we seek to re-engineer, repair and refactor.
Maintenance problems, such as these, all reside within language-rich problem domains.
It is therefore unsurprising that this area finds many applications of LLM-based techniques, as we review in this section.

\subsection{Debugging}
Kang et al.~\cite{kang2023preliminary} studied GPT-3.5's fault localisation ability,
and found that LLM could often identify the faulty method on the first try.
Wu et al.~\cite{wu2023large}
present a comprehensive investigation into the capability of GPT-3.5 and GPT-4 for fault localisation accuracy, stability, and explainability. 
The results demonstrate that GPT-4 achieves 
47\% higher fault localisation accuracy over the state-of-the-art, 
but the performance declines dramatically with a longer code context.

Feng and Chen~\cite{feng_prompting_2023} proposed AdbGPT, which reproduces Android bugs from bug reports through prompt engineering with ChatGPT.
With a dataset of 88 bug reports, AdbGPT was able to successfully reproduce 81\%, outperforming the baselines and ablations.
Joshi et al.~\cite{joshi_repair_2022} focused on multilingual debugging and proposed RING, which 
proposes a prompt-based strategy that conceptualizes repair as localization, transformation, and candidate ranking.

To address the data leakeage threat in fault localisation and program repair, Wu et al.~\cite{wu2023condefects} introduced ConDefects with
 1,254 Java bugs
and 1,625 Python bugs that were produced between
October 2021 and September 2023.
Researchers are allowed to select code samples based on their creation period, thereby allowing them to evaluate the effectiveness
of different LLMs according to their training data cut-off date.
In addition, there has been work on predicting bug severity with LLMs~\cite{mashhadi2023method}.

\subsection{Program Repair}
\spec{Repairing bugs has been a topic of much interest in the last decade in the software engineering research community. 
This section will explore open research questions and challenges for LLM-based repair approaches, including bug fixing, bug localization, root cause analysis, and bug remediation. }

Repairing bugs has been a topic of much interest for over a decade in the software engineering research community \cite{legoues:genprog,huang2023survey}, and has already found its way into initial industrial deployment \cite{ametal:sapfix}.

Much of the work on automated repair has used the generate-and-test approach widely adopted in the field of Genetic Improvement and readily applicable to LLM-based techniques.
As a result, LLMs are certain to have a positive impact on automated software repair, but there remain technical challenges in taming the hallucination problem and managing scalability, as we report in this section.

In order to achieve scalability, all generate-and-test approaches need to address the build time problem \cite{mh:apr22-keynote}.
LLM-based repair is no exception; the propensity to hallucinate makes it all the more important that the test phase can be executed regularly.
It is likely that using ReAct deployment models \cite{yao:react} will help to find efficient and effective engineering trade-offs.
When ReAct is applied to repair, the overall approach would alternate between the `Reason' phase (generating candidate fixes) and the `Action' phase (evaluating fixes through testing, which involves the build problem).

To address this issue, we can refer to the well-established literature on software repair \cite{legoues:cacm-survey,monperrus2018automatic}, grounded in over two decades of the development of search-based approaches to software engineering \cite{mhbj:manifesto,mhamyz:acm-surveys}.
This literature provides the  research community with a firm foundation of experience and expertise, making it very well-placed to develop LLM-based generate-and-test approaches to the problem.

Recent work on repair has started to use neural AI models, such as the seminal work of Tufano et al.~\cite{tufano:empirical}.
More recently, since 2022, there has been a rapid development of emergent embryonic research literature on LLM-based repair.
For example, Xia et al. \cite{xiatraining} proposed AlphaRepair.
It redefines the APR problem as a cloze (or infilling) task, where the LLMs are leveraged to directly fill-in correct code based on the bi-directional context of the potential buggy code portions. AlphaRepair also demonstrates for the first time that LLMs can outperform all prior APR techniques. 

They further conducted an empirical study~\cite{xia:practical} on nine LLMs across five datasets in three different languages.
Their findings not only affirmed the superiority of LLM-based APR (especially the cloze-style approach) but also offered a number of practical guidelines.
Wei et al.~\cite{wei2023copiloting} synthesize a patch through the
interaction between an LLM and a Completion Engine, and found that the approach surpasses the best-performing baseline by 14 and 16 bugs fixed.

Program repair naturally fits a conversational model of prompt engineering. 
Xia et al.~\cite{xia_conversational_2023} proposed 
conversational APR, which alternates between patch generation and validation in a conversational manner.
Their evaluation on ten LLMs demonstrated that their approach has superiority in both effectiveness and efficiency.

They further proposed ChatRepair~\cite{xia_keep_2023},
showing that the conversational approach fixes 162 out of 337 bugs for only \$0.42 per bug, thereby also addressing potential concerns about the computational resources required.
Chen et al.~\cite{chen_teaching_2023} 
introduced SELF-DEBUGGING, which teaches an LLM to debug its predicted code via few-shot learning,
SELF-DEBUGGING reports  baseline accuracy improvements of up to 12\%.

Studies have also reported results for particular classes of bugs, for example,
Pearce et al.~\cite{pearce:examining} reported repair results from five commercial LLMs on  security bugs,
Charalambous et al.~\cite{charalambous2023new} combined ChatGPT with 
 with
formal verification strategies to verify and automatically repair
software vulnerabilities.
Cao et al.~\cite{cao_study_2023} report ChatGPT results for Deep Learning (DL) program repair. 


Repair does not always start with an existing failing test case, but can start with a natural language description of a failure in production.
Automation opens the door to faster responses to user-generated bug reports.
This is a route to repair  that has also been explored for LLMs in the work of Fakhoury et al. \cite{fakhoury_towards_2023},  who generated functionally correct code edits from natural language issue 
descriptions. 
They propose Defects4J-Nl2fix, a dataset of 283 Java programs from the Defects4J dataset with high-level descriptions of bug fixes. The state-of-the-art LLMs evaluated on this 
benchmark achieve up to 21\% Top-1 and 36\% Top-5 accuracy.

Automated repair can also reduce the burden on engineers, managing DevOps-style on-call for production systems. For example, Ahmed et al.~\cite{ahmed_recommending_2023} studied the use of LLM-based 
root causing and remediation of 40,000 incidents on Microsoft cloud services. The authors evaluated multiple LLMs using semantic and lexical metrics in zero-shot, fine-tuned, and multitask settings, 
showing that fine-tuning significantly improves  incident response effectiveness.

The ability to perform fine-tuning for a specific task or domain can significantly improve the model performance in program repair. 
Jiang et al.~\cite{jiang:impact}  empirically evaluated the performance of 10 different Code Language Models (CLMs) and 4 fault benchmarks, and showed that repair-specific fine-tuning could significantly improve success rates. 
On average, the 10 CLMs already successfully repaired 72\% more faults than state-of-the-art DL-based APR techniques. After fine-tuning, the number increased to 160\%.
Jin et al.~\cite{jin_inferfix_2023} proposed InferFix, which contains a LLM (Codex Cushman) finetuned on supervised bug-fix data.
InferFix achieves a 76\% Top-1 repair accuracy on Java, and over 65\% on C\# using the InferredBugs dataset.
Berabi et al.~\cite{berabi:TFix} introduced TFix, a T5 model fine-tuned on bug-fixing data, reporting that it outperformed existing learning-based approaches.
Xia et al.~\cite{xia2023revisiting} combines LLM fine-tuning and prompting to automate the plastic surgery hypothesis and demonstrated that their approach 
fixes 89 and 44 bugs (outperforming the baseline by 15 and 8).

LLMs can also help to explain the patches that they generate.
Kang et al.~\cite{kang2023explainable} proposed AutoSD to provide debugging explanation with LLMs to help developers judge the correctness of patches.  
They found that AutoSD produced comparable results to existing baselines with high-quality repair explanations.
Sobania~\cite{Sobania:2023:SSBSE} studied the capability of
GPT 3.5 in
explaining the patches generated a search-based repair tool, ARJA-e, on 30 bugs from Defects4J.
84\% of the LLM explanations are found to be correct.

\subsection{Performance Improvement}
\label{sec:perf}
\spec{changes to code to improve performance is a particularly interesting  class of small code modifications. The section will explore challenges in tackling performance improvement (execution speed, memory footprint size, latency etc.) using LLMs}

Since the inception of computer programming, the paramount importance of performance optimisation has been recognised.
Indeed, performance optimisation is even mentioned by Ada Lovelace in her nineteenth-century notes on the analytical engine \cite{lovelace:sketch}.
Much initial practical deployment of optimisation took place in compiler development, through work on optimising compilers \cite{aho:dragon}.
This is the bedrock on which current practical and efficient computation rests, but it is necessarily a one-size-fits-all approach; widely applicable due to its generality, yet suboptimal for bespoke problem domains for the same reason.
There has, therefore, also been much work on specific source-to-source transformations to improve optimisation, dating back to the 1970s \cite{darlington:system,partsch:cip}.

For a long time, the focus of this work was on finding suitable sets of meaning-preserving transformations, the motivation being that a correct program can be transformed into a more efficient version 
of itself, while retaining its correctness. However, more recently, research on program synthesis took a different turn: Inspired by Genetic Programming \cite{koza:genetic}, and early results from 
Automated Program Repair \cite{legoues:genprog,perkins:auotfix}, it considered a wider set of transformations in an approach that has come to be known as `Genetic Improvement' \cite{Petke:gisurvey,mhetal:seams14-keynote}.

The wider set of transformations may produce incorrect code, but automated testing can filter these, to ensure 
sufficient faithfulness to the intended semantics. 
Furthermore, the freedom to treat existing code as a kind of `genetic material' produced dramatic improvements in non-functional properties, such as 
execution time, memory and power consumption (e.g., 70x speed up of a non-trivial gene sequencing system \cite{blmh:tec1}).

Although the potential for artificial intelligence techniques, such as evolutionary algorithms, to improve performance has been well studied, researchers have only just begun to consider the potential 
for LLM-based performance improvement. In the work by Madaan et al. \cite{madaan_learning_2023}, the authors use CODEGEN and CodeX to suggest functionally correct, Performance-Improving Edits (PIEs), 
improving execution time of Python and C++ (already pre-optimised with the maximally optimising compiler option {\tt -O3}). Similarly, Garg et al.
~\cite{garg2022deepdev} proposed DeepDev-PERF, a
performance improvement suggestion approach
for C\# applications that.
DeepDev-PERF took the
English-pretrained BART-large model and further
pretrained it on Source code.
Kang and Yoo~\cite{kang2023objectivetailored} proposed the use of LLMs to suggest objective-specific mutation operators for genetic improvement, and provided demonstrations on improving efficiency and decreasing memory consumption. 
Garg et al.~\cite{garg2023rapgen} proposed RAPGen, which generates zero-shot prompts for LLMs to improve performance.
The prompts are generated via retrieving a prompt instruction from
a pre-constructed knowledge base of previous performance 
improvements. 
Chen et al.~\cite{chen2023supersonic} used GPT models as baselines for their source code optimisation method, Supersonic, and found that Supersonic improves running
time for 26.0\% of the programs, compared to only 12.0\% for
GPT-3.5-Turbo and 4.0\% for GPT-4.

Cummins et al.~\cite{cummins:large} focused on the performance of compilers and 
presented results on  LLMs for optimising compiler instructions. 
Their results  demonstrate that a relatively small (7B-parameter) LLM, trained to generate instruction counts and optimized compiler LLVM code, can generate 3\% improvements in reducing compiler instruction counts, 
outperforming the state-of-the-art. 
Their results are also promising in terms of correctness, with 91\% compilable and 70\%  functionally correct wrt the original compiler output.

\begin{figure}[ht]
\includegraphics[width=90mm]{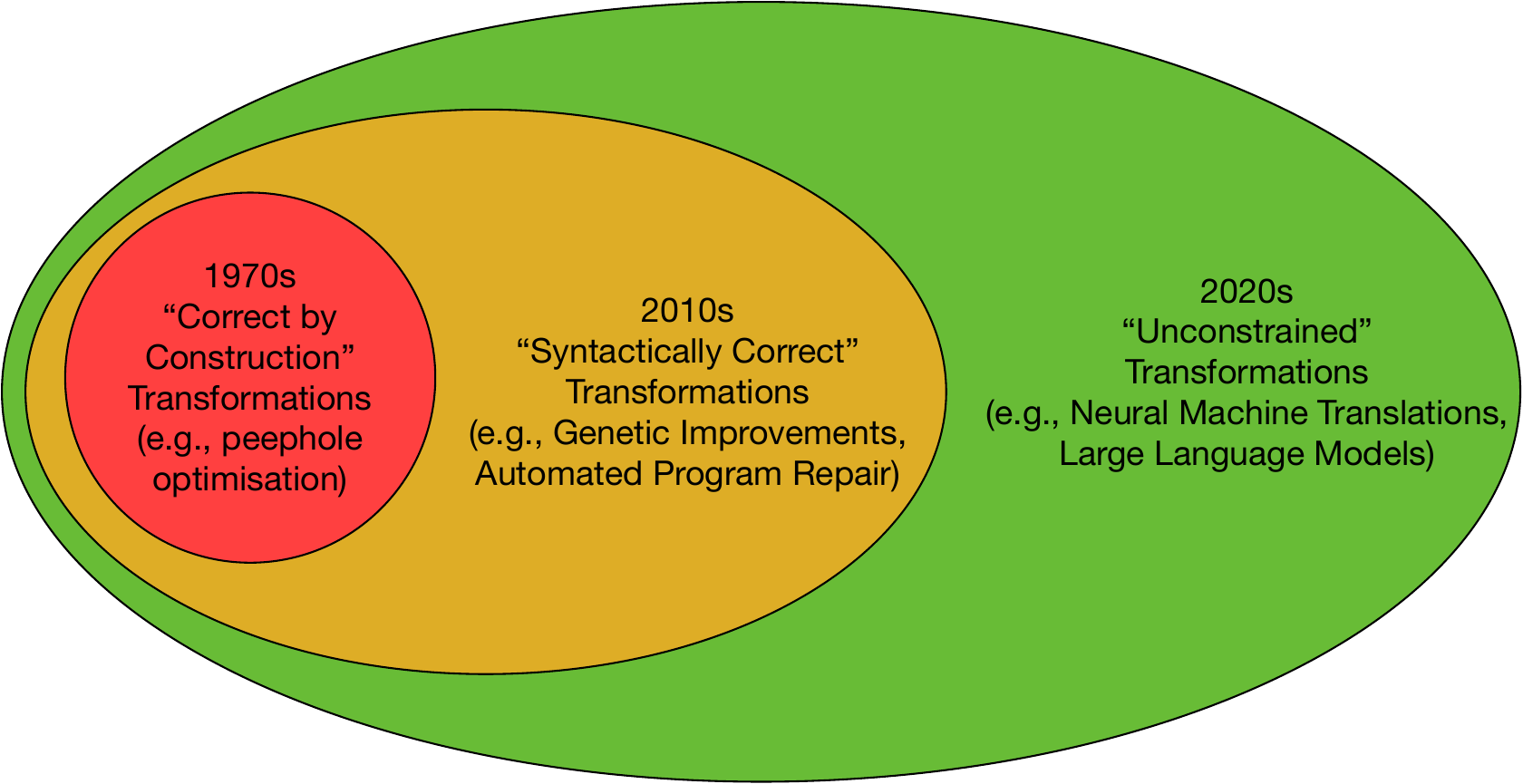}
\caption{The Widening Scope of Program Transformation\label{fig:onion}}
\end{figure}

Over a period of some 50 years, the software engineering community has evolved its conception of what it means to transform an existing software system into an equivalent system that improves performance while retaining functional behaviour. 
In the 1970s, the strongest concern was correctness, so transformation palettes were defined to 
consist solely of transformation steps that were (functionally) correct by construction.

However, by 2010 the community was already exploring the application of considerably more relaxed notions of equivalence that merely retain sufficient operational faithfulness to the behaviour of the original. 
The tight semantic straitjacket of the 1970s was thereby considerably relaxed to allow transformations that might even {\em fail} some test cases. 
During the same period,  operational performance became increasingly important.
A key underlying principle of this research agenda is that no overall software system can be deemed functionally correct, when it is executed on a system in which inefficiency has left insufficient remaining resources.
This principle applies even in the (comparatively rare) cases where the software has been fully proven to be functionally correct.
As the more pithy slogan has it:

\begin{quote}
``There is nothing correct about a flat battery''~\cite{Petke:gisurvey}.
\end{quote}

This evolution of the community's approach to code transformation and synthesis is depicted in Figure~\ref{fig:onion} (red and yellow regions).

In the context of this increasing relaxation of semantic constraints, we can view LLM-based code optimisation as a further development of this overall direction of travel: 
Code optimised by LLMs may not be even {\em syntactically} correct, let alone semantically correct (depicted by the green region of Figure~\ref{fig:onion}). 

Despite these correctness challenges, inherent in LLM-Based SE, there is a large pool of training data, and LLMs have a  propensity to exhibit emergent behaviour.
These observations combine to yield surprising results that, although not guaranteed to be correct, can potentially dramatically change performance characteristics in useful ways.

Of course, as we increasingly allow more permissive transformation pallets in the hope of optimising multiple non-functional properties, we simultaneously place far greater reliance upon the ability of testing to provide reassurance of functional faithfulness. 
Testing is also vital to check for regressions in those non-functional properties that are not targeted by the improvement process.
As a result, software testing in general (and automated high coverage test generation in particular), will become ever more important.

\subsection{Clone Detection and Re-use}

There has been much previous work on managed software reuse \cite{krueger:reuse} in order to extract value and avoid duplication, a topic also tackled using LLMs
\cite{huang:PCR}.
Software typically contains large numbers of clones, arising from ad hoc re-use, resulting in much work on automated clone detection \cite{zakeri:systematic}, a topic for which fuzz-based fine-tuned LLMs have also been applied
~\cite{zhao2023understanding}.

\subsection{Refactoring}
\label{sec:refactoring}
\spec{when we refactor code we expect its behaviour to remain the same. 
This is particularly attractive because it means that we can simply rely on regression testing which already has an Oracle. 
The section will outline challenges and opportunities for tackling refactoring using LLMs}

When we refactor code, we generally expect its behaviour to remain unchanged. This is particularly attractive for automated approaches (such as search-based refactoring \cite{mariani:systematic-refactoring}) because it means that we can simply rely on the Automated Regression Oracle. This `automatable oracle for free' advantage is significant and will also apply to LLM-based refactoring.



Poldrack et al.~\cite{poldrack_ai-assisted_2023} show that GPT-4 refactoring of existing code can significantly improve code quality according to long-established 
structural metrics such as Halstead \cite{halstead:complexity} and McCabe \cite{mccabe:complexity} complexity.
Noever and Williams~\cite{noever_chatbots_nodate} emphasize the value of AI-driven code assistants in refactoring legacy code and simplifying the explanation or functionality of high-value repositories. 

\subsection{Open Problems in Maintenance and Evolution}
Since so many of the subdomains of software maintenance and evolution are concerned with existing legacy system source code, we can expect rapid growth in the application of LLMs. 
This section outlines some existing open problems in this nascent sub-area of research.

\subsubsection{Open Problems in Performance Improvement}
Much more work is needed on the development of LLM-based techniques for automatically finding performance improvements. As with Genetic Improvement, these need not be confined merely to execution time, but can also consider other non-functional attributes such as power consumption \cite{bbetal:approximate,li:making,white:searching} and memory footprint~\cite{fan:gecco15}
as well as multi-objective, trade-offs between sets of non-functional properties~\cite{mh:ase-keynote}. We expect more work on Genetic Improvement-style LLM-based code optimisation techniques, with the potential for many dramatic advances and breakthroughs.

\subsubsection{Open Problems in Refactoring}
By definition, refactoring does not change semantics, so LLM-based refactoring can rely on the Automated Regression Oracle. It is therefore surprising that there is not already more work on LLM-based refactoring. In this subsection, we outline possible directions.


Design patterns have played a critical role in practical software engineering for three decades~\cite{gamma:design}. LLMs may help engineers to refactor existing code to use design patterns, while providing developer-friendly explanations and documentation. 

Refactoring also becomes necessary whenever new technologies emerge.
For example, when an API is updated or a new  API becomes available.
Although they can be (sometimes automatically \cite{mketal:evaluating-repair}) repaired, API misuse remains a common source of software engineering bugs.
Automating the process of refactoring for new APIs is less challenging than other code transformations, because of the presence of the Automated Regression Oracle.

Finally, the few-shot learning capabilities of LLMs may enable more bespoke refactoring. 
The emergent work on LLM-based refactoring has focused on global refactoring according to well-known refactoring patterns.
However, programmers often have project-specific refactoring requirements.
Up to a third of software engineering effort is spent on largely repetitive, tedious, and potentially error-prone refactoring activities that implement these project-specific refactoring needs.
The few-shot learning potential of LLMs may automatically generalise from specific examples, automating what we call `bespoke' refactoring.
More work is needed to develop techniques for reliable few-shot-learnt bespoke refactorings.

\section{Documentation generation}
\spec{Software is not just code. It includes other artefacts, such as documentation. However, whereas code can be executed (thereby giving a  ground truth), documentation typically cannot. This raises challenges for the research community to work out techniques that can help us generate assured documentation; misleading documentation is worse than no documentation at all. 
This section will outline these challenges in existing work in this area }

Most of the work on LLM-based software engineering has focused on the generation of code, but there is also considerable potential for LLM-based documentation generation.

Sun et al.~\cite{sun_automatic_2023} explored how ChatGPT performs on code summarisation of Python code. They used CSN-Python and compared ChatGPT with NCS, CodeBERT, and CodeT5.
They adopted three widely-used metrics: BLEU, METEOR, and ROUGE-L. Surprisingly, the results show that ChatGPT’s performance is significantly worse than the baseline models in terms of BLEU and ROUGE-L.

Ahmed et al.~\cite{ahmed_improving_2023} conducted prompt engineering for code summarisation on GPT-3.5, while
Geng et al.~\cite{geng_empirical_2023} performed experiments on two  Java language datasets, Funcom and TLC, using Codex:  to generate 
\emph{multiple-intent} comments.
Gent et al.~\cite{Geng:summarizers} demonstrate that pre-trained LLMs already have sufficient context to generate multiple different code summaries from different technical perspectives.

\subsection{Open Problems in Documentation Generation and Code Summarization}

Many existing code summarization techniques are retrieval-based: the given code is represented in a vector format using a neural representation, which is subsequently used to retrieve the most 
relevant textual summarization from the corpus. 

There is a clear limitation to this approach due to the fact that the set of summaries that can be generated are constrained by the training corpus. 
LLMs may 
enable automated code summarization that is not restricted to this training corpus, assisted by their natural language processing capabilities. 

While this may result in richer and more semantically 
relevant summaries, we also note that existing evaluation metrics are often lexical in nature, hindering our ability to compare and evaluate richer summaries generated by LLMs~\cite{sun_automatic_2023}.
Recent advances in ReAct-based approaches \cite{yao:react} may open up other avenues for greater assurance in the documentation generated, even when it cannot be executed.

\section{Software Analytics and Repository Mining}
\spec{there is a well-established field on software analytics; essentially how to yield insight for human engineers from existing software artefacts. The section will consider ways in which LLM-based approaches can tackle existing known problems in software analytics, and the new challenges and opportunities it will pose for analytics of LLM-generated software.
The section also considered the potential for LLM-generated software to be highly understandable and human-maintainable. That is, by incorporating software analytics, self-documentation and traceability into the generated code at the time it is generated.
In this way, we can hope that software engineered by machines will smoothly interoperate with software engineered by humans. 
Arguably, it will potentially more smoothly inter-operate than different pieces of software written by different human engineers.}

There is a well-established field of software analytics;  how to yield insight for human engineers from existing software artefacts \cite{menzies:sowhat}.
The large amount of software artefact information publicly available online has stimulated the growth of scientific insights gained by Mining Software Repositories (MSR) \cite{hassan:road,martin:tse-survey}.
While MSR tends to focus on scientific research insights from such mining, software analytics tends to focus on opportunities for organisations to gain insight from the analysis of their own repositories, which can also benefit AI understandability~\cite{menzies:software}.

Hitherto, in both cases, much of the collection, curation and analysis
of data has relied upon labour-intensive human analysis.
We found no work on the use of LLMs to support this activity.
Nevertheless, because many LLMs have already ingested this software artefact data, and are capable of providing reasoning and insight, it seems natural to expect them to play a significant role.

For example,  LLMs may identify interesting new MSR research questions, based on their ability to ingest large amounts of data, including 
research questions and hypotheses that have previously proved interesting to researchers.
They may also assist with traceability, which 
software engineers have great difficulty maintaining \cite{spanoudakis:software,cleland:software}.

\section{Human Computer Interaction}
\spec{finding productive interfaces between human engineers and software infrastructure has remained a recurring theme throughout the lifetime of the development of software engineering over the last five decades. This section will focus exclusively on engineer-machine interaction, and will not discuss user-machine interaction. This will ensure that the material in the section remains entirely uncontroversial, and focused on technical software engineering aspects.}

Finding productive interfaces between human engineers and software infrastructure has remained a recurring theme throughout the lifetime of the development of software engineering \cite{lebeuf:software,lehman:essential}, dating back to the inception of the discipline in the 1960s \cite{hamilton:universal}. 

We found evidence of many interesting research questions. 
For example, Vaithilingam et al. \cite{vaithilingam_expectation_2022} reported on the  difficulties 24 participants had in understanding, editing, and debugging the Copilot-generated code,
while 
Feldt et al.~\cite{feldt:autonomous} proposed a hierarchy of design architecture for LLM-
based software testing agents.
Liang et al.~\cite{Liang:large-scale} surveyed 410 practising software engineers, finding widespread use of LLMs to facilitate low-level programming tasks, but also resistance to using LLMs for more design-level software engineering activities.
Feng et al. \cite{feng_investigating_nodate}  collected 316K tweets and 3.2K Reddit posts about ChatGPT’s code generation to understand social media's attitudes toward AI-assisted coding tools.

They found that fear is the dominant emotion associated with ChatGPT’s code generation, overshadowing other emotions such as happiness and surprise.
Ahmad et 
al.~\cite{ahmad_towards_2023} explore the way in which a novice software architect could interact with ChatGPT.

\section{Software Engineering Process}
Software engineering concerns, not only software products, but also the process by which they are constructed~\cite{Pressman:92}.
Previous research on software assistants \cite{kbetal:esem21-keynote,lebeuf:software,cornes:coq,gallagher:surgeon,ward:maintainers} is clearly of particular relevance to LLM-based 
software engineering, a topic
some authors have already started to consider. 
For example, Ross et al.~\cite{ross_programmers_2023}, introduced an LLM-based programmers' assistant, evaluating its deployment with 42 
participants while  Tian et al.~\cite{tian_is_2023} highlighted the attention span limitations of ChatGPT.

\section{Software Engineering Education}
\label{sec:education}
Teachers have expressed concern at the difficulties of identifying cases where students have relied on 
LLMs to construct their assignments~\cite{meckler:alert}, while other authors have argued that the long-term impact of LLMs on education will be beneficial~\cite{heaven:chat}.
However, our present focus rests more narrowly on the specific impact of LLMs on the field of {\em software engineering} education, where the current literature focuses on LLM-based tutorial support.

For example, 
Jalil et al.~\cite{jalil_chatgpt_2023} explored opportunities for (and issues with)  ChatGPT in software testing education.
Savelka et al.~\cite{savelka_large_2023}  analysed the effectiveness of 
three models in answering multiple-choice questions from introductory and intermediate programming courses at the postsecondary level. 
Several other authors ~\cite{sarsa_automatic_2022,macneil_experiences_2023,noever_chatbots_nodate} explored the capabilities of CodeX for generating programming exercises and code explanations.
Their general finding was that the majority of the  generated content was 
novel, sensible, and useful (see also 
Section~\ref{sec:explanability}).

\section{Crosscutting Open Research Topics}
\label{sec:opentopics}
A number of patterns emerge from the embryonic literature on LLM-based software engineering.
In this section, we outline those that raise open research questions that cut across all software engineering applications

\subsection{Building and Tuning LLMs for SE}

Most of the previous work has treated LLMs as atomic components, with a focus on incorporating these in wider software engineering workflows. 
While there have been attempts to tailor the behaviour, these have tended to focus on prompt engineering, with a few examples of fine-tuning.

A more challenging but potentially impactful problem lies in training and fine-tuning models, specifically for software engineering tasks. 
Ding et al.~\cite{Ding:TRACED}  train a BERT-like LLM with execution inputs and dynamic execution traces. 
They show how this dynamic information improves (up to 25\%)  the accuracy of the model for downstream software engineering predictive tasks: vulnerability and clone detection and coverage prediction (full execution path and branch coverage).

More work is needed on new forms of LLMs, specifically tailored for software engineering that take advantage of software's unique properties and distinguish it from natural language. Dynamic information is one such key differentiator currently missing from most of the work.
We expect the next generation of SE-specific LLMs to address this.

An important aspect of building and training LLMs is their energy consumption.
LLM capabilities  
have been associated with their size~\cite{kaplan2020scaling}, resulting in rapid growth of 
model size~\cite{chowdhery2022palm-short,rae2022scaling-short}. 
The training and developing of larger models may have direct environmental impact~\cite{Henderson2020}. While it has been suggested that the model performance depends not only on model size but also on the volume of training data~\cite{hoffmann2022training}, the question of the right model size required to achieve the desired performance remains unclear. 

Lighter models may also widen adoption, thereby leading to enhanced deployability. 
Recently, techniques such as low-rank adaptation (lora)~\cite{hu2021lora} and model quantization~\cite{polino2018model} have shown potential, but they remain to be empirically evaluated with respect to specific applications.

\subsection{The Need for Dynamic Adaptive Prompt Engineering  and Parameter Tuning}
Initial work on prompt engineering has demonstrated its potential to considerably improve the software engineering artefacts generated by LLMs.
However, as already found  \cite{doderlein_piloting_2023}, the results are highly problem-specific,  so a one-size-fits-all approach is unrealistic.
Furthermore, very few papers report model parameter settings, yet we know that many of these, such as the temperature setting, play a crucial role in determining the nature  of the generated LLM output.

As an immediate starting point, it is imperative that authors make a point of conspicuously reporting these parameter settings to support replication. 
However, we also need more research on dynamic adaptive prompt engineering and model parameter tuning.
This research agenda may draw inspiration from existing work on parameter tuning for other dynamic adaptive tasks, such as fuzzing \cite{manes:fuzzing}.
Dynamic prompt optimisation may also exploit techniques associated with SBSE \cite{mhamyz:acm-surveys}, reformulating prompt optimisation as a multi-objective computational search process.

\subsection{Hybridisation}
LLMs are seldom most effective when used in isolation, but can be highly effective as part of an overall SE process.
More work is needed to understand the design patterns for SE workflows into which LLMs can safely, efficiently and effectively reside.
We believe that existing SE theory and practice associated with generate-and-test approaches, such as Automated Repair and Genetic Improvement, are already highly amenable to LLMs.

We expect to see much more work incorporating LLMs into these existing software engineering frameworks.
However, more work is required to tailor and extend these frameworks,  to best take advantage of the opportunities offered by LLM-based software engineering.

In particular, we expect to see a rapid development of work on static and dynamic analyses for prompt engineering and post-processing of LLM responses.
We also expect to see hybrid software engineering processes, adapting Continuous Integration pipelines to incorporate LLMs.

\subsection{Harnessing Hallucination}
While hallucination has widely been regarded as a problem, as reported in this survey, it may also prove to provide benefits when applied to software engineering domains.
LLM hallucinations are seldom entirely {\em random} incorrect responses.
Rather, because of their inherent statistical properties, 
they would be better characterised as
`plausible futures', and this may often make them useful when set in the right context.

Hallucination can be repurposed to provide potentially {\em useful suggestions} for software enhancement.
For example, 
when hallucinating a test case, the LLM may be repurposed to suggest new features, while
a hallucinated code summarisation might indicate potential for (human) code misunderstanding; if the LLM `misunderstood' the code, might not a human also misunderstand it? 
When the LLM hallucinates an non-existent API, it may be repurposed as a way to {\em suggest} refactoring to simplify or extend existing APIs.
More work is needed to exploit this positive potential, and to harness hallucination for software improvement.

\subsection{Robust, Reliable, and Stable Evaluation}
Hort et al. \cite{hort:exploratory} conducted a review of 293 papers on LLMs for code generation, to determine the degree to which sufficient information was shared to support replication.
They found that only 33\% shared source code and 27\% shared trained artefacts.  
They also evaluated the papers from the perspective of energy consumption, assessing the degree to which it was possible for an independent researcher to assess the energy consumed during training. 
They report that approximately 38\%  (30 out of 79 publications which involve model training) shared
sufficient information to estimate their energy consumption during training. 

Further evidence that there may be a growing issue with scientific evaluation quality in the literature on LLM-Based Software Engineering can be found in the survey of LLM-Based Testing by Wang et al. \cite{wang:llm-test-survey}.
In their survey, they filtered an initial pool of papers on LLM-Based Testing to remove those that did not meet standard evaluation quality constraints.
These constraints required papers to include a  
clear,
defined,
repeatable
evaluation methodology that includes a control or baseline against which to measure effectiveness.
This filtration criterion removed more than 90\% of the papers that initially met keyword search criteria.

As these analyses of the literature demonstrate, 
more work is clearly needed to establish firm scientific foundations for the emerging discipline of LLM-based Software Engineering.
Such foundations may draw on existing foundations for Empirical Software Engineering in general and, more specifically, on AI-based Software Engineering, such as SBSE (where there is a natural similarity~\cite{tang:chatgpt,lemieux_codamosa_nodate}).

Nevertheless, LLMs have their own unique properties, such as the ability to provide explanations,  which will require domain-specific theoretical and empirical scientific foundations. 

LLMs inherently exhibit non-deterministic behaviour.
Researchers need to carefully design their experiments, configure their LLMs 
(e.g., evaluating the effects of different distribution sampling strategies), 
and take into account non-determinism when drawing
their conclusions on LLMs.
The SBSE literature provides advice on the inferential statistics required to support such evaluation~\cite{arcuri:practical,mhetal:sbse-tutorial}.

We will witness a rapid growth in the number and diversity of language models for software engineers in the coming years.
Both practitioners and practising software engineers will need reliable, efficient and comprehensive benchmarking systems.
Benchmarking platforms such as TESTPILOT \cite{schafer_adaptive_2023} and platforms such as Papers With Code (\url{https://paperswithcode.com/sota/code-generation-on-humaneval/}) 
will become increasingly important.

As well as generic scientific foundations, benchmarks and evaluation platforms, we also expect to see longitudinal studies of developer behaviour when programming with LLM assistance, so that we can understand the programmer-LLM interaction better and design more effective use case scenarios.

\subsection{Thorough Testing}
The problem of hallucination has already been widely studied. 
It will continue to be a topic of great interest, both within the software engineering community and in the wider computer science community.
While it is likely great progress will be made, the inherent risk of hallucination is unlikely to be completely eradicated, since it is as germane to the LLM technology, as it is to human intelligence.
Fortunately, over more than six decades, software engineers have developed robust automated verification and testing technologies that help to reduce the impact of human mistakes. We expect that such technologies will also  carry over to artificial intelligence mistakes.

\subsection{Handling Longer Textual Inputs}
The performance of LLMs on large-sized input prompts is likely to be a topic of great interest in the artificial intelligence community~\cite{shaham_scrolls_2022}. 
Advances in this area will have a strong impact on software engineering, 
because of the considerable size of software systems and the consequent opportunities that additionally open when larger prompts are to be effectively handled.

\subsection{Less Well-covered Subdomains of Software Engineering}

As our survey reveals, some subdomains of software engineering are notably under-represented in the literature; some surprisingly so.
For example,
Requirements Engineering and Design (Section~\ref{sec:design}),
and
Refactoring (Section~\ref{sec:refactoring}) enjoy very little coverage, yet they are surely ripe for consideration, since they rely heavily upon linguistic forms of analysis and the recognition and prediction of patterns.

\bibliographystyle{IEEEtran} 

\bibliography{Citation} 

\end{document}